\documentclass[journal]{IEEEtran}
\usepackage{cite}
\usepackage[backref]{hyperref}
\usepackage{amsmath}

\ifCLASSINFOpdf
   \usepackage[pdftex]{graphicx}
   \graphicspath{{../pdf/}{../jpeg/}}
   \DeclareGraphicsExtensions{.pdf,.jpeg,.png}
\else
   \usepackage[dvips]{graphicx}
   \graphicspath{{../eps/}}
   \DeclareGraphicsExtensions{.eps}
\fi

\usepackage{caption}

\usepackage{multirow}
\usepackage{float}

\usepackage{graphicx}

\ifCLASSOPTIONcompsoc
  \usepackage[caption=false,font=normalsize,labelfont=sf,textfont=sf]{subfig}
\else
  \usepackage[caption=false,font=footnotesize]{subfig}
\fi

\begin{document}
\title{A Radar Signal Deinterleaving Method Based on Semantic Segmentation with Neural Network}

\author{Wang~Chao,
        Sun~Liting,
        Liu~Zhangmeng,
        and~Huang~Zhitao% <-this % stops a space

\thanks{Wang Chao is with College of Electronic Science and Technology and College of Electronic Engineering, National University of
Defense Technology, Changsha 410003, China, E-mail: (wangchaoben@126.com)}
\thanks{Sun Liting, Liu Zhangmeng, and Huang Zhitao are with College of Electronic Science and Technology, National University of
Defense Technology, Changsha 410003, China, E-mail: (slt2009@yeah.net; liuzhangmeng@nudt.edu.cn; huangzhitao@nudt.edu.cn)}

\thanks{Corresponding author: Liu Zhangmeng}
}

%\thanks{M. Shell was with the Department
%of Electrical and Computer Engineering, Georgia Institute of Technology, Atlanta,
%GA, 30332 USA e-mail: (see http://www.michaelshell.org/contact.html).}% <-this % stops a space
%\thanks{J. Doe and J. Doe are with Anonymous University.}% <-this % stops a space
%\thanks{Manuscript received April 19, 2005; revised August 26, 2015.}}

\maketitle
\begin{abstract}
Radar signal deinterleaving is an important part of electronic reconnaissance. This study proposes a new radar signal deinterleaving method based on semantic segmentation, which we call "semantic segmentation deinterleaving" (SSD). We select representative sequence modeling neural network (NN) architectures and input the difference of time of arrival of the pulse stream into them. According to semantics contained in different radar signal types, each pulse in the pulse stream is marked according to the category of semantics contained, and radar signals are deinterleaved. Compared to the traditional deinterleaving method, the SSD method can adapt to complex pulse repetition interval (PRI) modulation environments without searching the PRI or PRI period.  Multiple rounds of search and merging operation are not required for radar signals with multiple pulses in a period. Compared to other deinterleaving methods based on NNs, the SSD method does not need to digitize the data and train a network for each target type. The SSD method also does not need to iterate input and output data. The proposed method has high robustness to pulse loss and noise pulses. This research also shows that recurrent NNs still have more advantages than convolutional NNs in this sequence modeling task.

\end{abstract}


\IEEEpeerreviewmaketitle

\section{Introduction}
\begin{figure}[!t]
\centering
\includegraphics[width=2.5in]{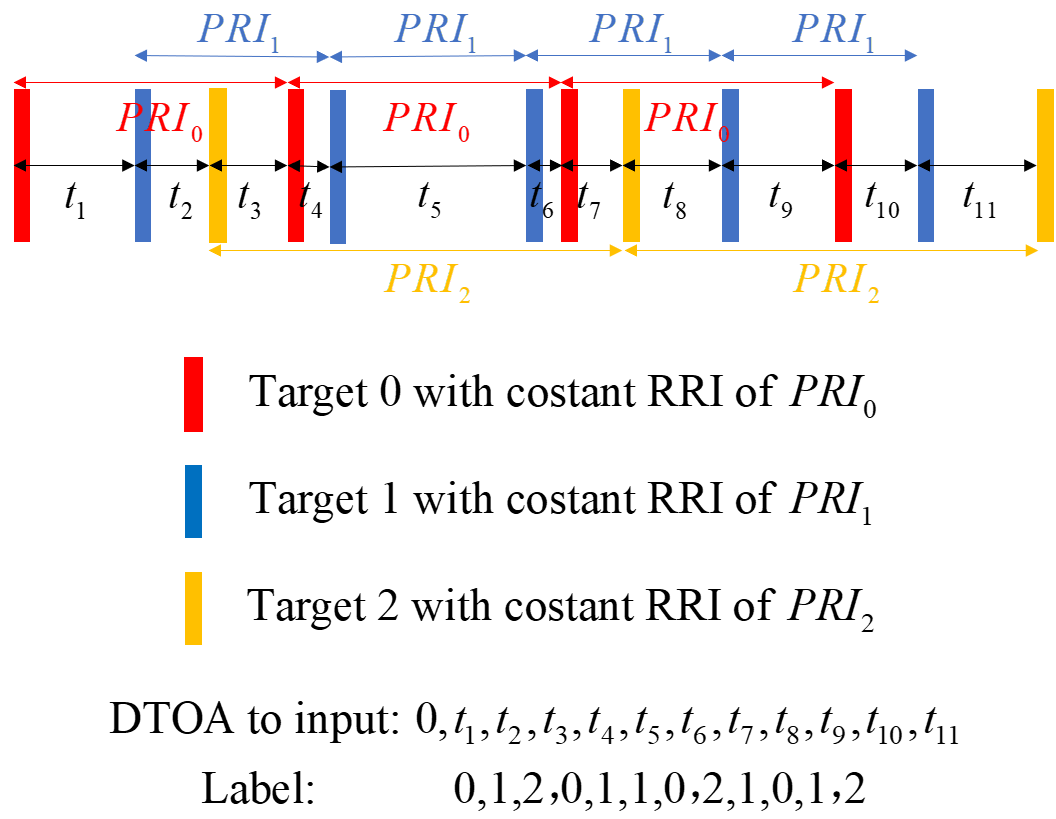}
\caption{Input and label of an interleaved pulse stream.}
\label{interleaved}
\end{figure}

In electronic warfare, to obtain information about a target radar, it is necessary to use electronic reconnaissance equipment for  reconnaissance and interception of the corresponding target radar signal. In an actual electromagnetic environment, there are often other electromagnetic signals beside the target radar signal. In that case, data collected by electronic reconnaissance equipment may contain information from different targets, and the intercepted pulse stream may also contain interleaved pulses from different radiation sources, as shown in Fig. \ref{interleaved}. Full pulse data are pulse description words for each pulse output in chronological order in electronic reconnaissance equipment. Pulse description words include the time of arrival (TOA), direction of arrival (DOA), pulse width (PW), radio frequency (RF), pulse amplitude (PA), and other information about each pulse. Radar signal deinterleaving, which is an important part of electronic reconnaissance, is essentially to deinterleave the interleaved pulse description words belonging to different radiation sources in the full pulse data.

Semantic segmentation is an important task in image processing. It uses feature information about different categories of targets, namely semantics, to mark each pixel in an image according to the category of the target the pixel belonged to. This method can segment different target types in an image and is also known as dense prediction \cite{2018-deeplabv3+}. In recent years, neural networks (NNs) have become the most important tool and a topic of active research of image semantic segmentation.
 
\begin{figure}[!t]
\centering
\includegraphics[width=2.0in]{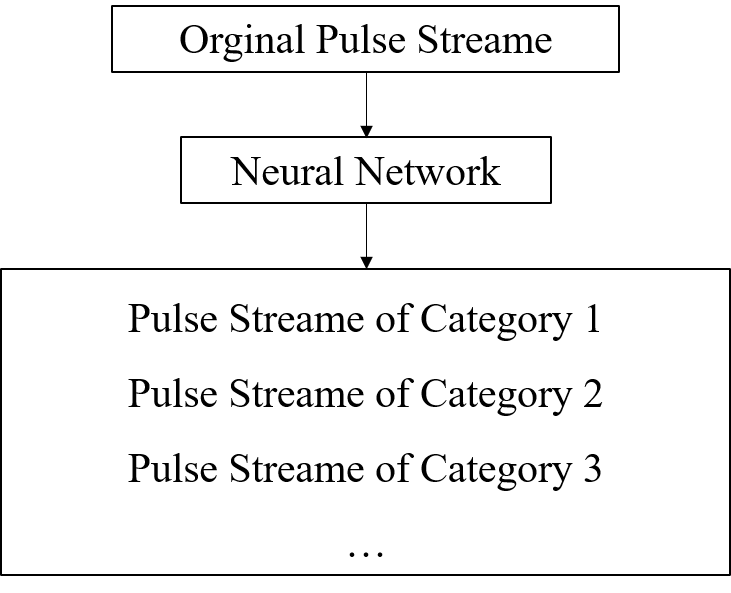}
\caption{Deinterleaving process of the SSD method.}
\label{SSDmethod}
\end{figure}

In this paper, a new radar signal deinterleaving method based on semantic segmentation is applied to radar signal deinterleaving, which we call “semantic segmentation deinterleaving” (SSD). According to feature information about different radar signal types, each pulse of interleaved pulse stream is marked according to the category of the target using NNs to deinterleave different radar signal types in a pulse stream. This method can realize signal deinterleaving of multiple radar targets through one network and one step, as shown in Fig. \ref{SSDmethod}. It has significant advantages over other methods.

Compared to the traditional methods \cite{1993-Wiley}, 
\cite{1989-CDIF}, 
\cite{1992-SDIF}, 
\cite{2000-PRITran}, 
\cite{1999-Spectrum-Estimation-of-Interleaved-Pulse-Trains}, 
\cite{2002-New-Matrix-Method}, 
\cite{1998-A-Novel-Pulse-TOA-Analysis-Technique}, 
\cite{1994-Kalman}, 
\cite{1998-Kalman}, 
\cite{1999-Kalman}, 
\cite{2010-Kalman-MHT},  
\cite{2005-HiddenMarkov}, 
\cite{2019-Improved-Deinterleaving-on-Correlation}, the proposed method does not need to search the pulse repetition interval (PRI) or PRI period and can adapt to complex PRI modulation. When there are multiple pulses in a period, there is no need to batch the results of multiple rounds of search.  Compared to other deinterleaving methods based on NNs and automata \cite{2020-lzm-Classification-Denoising}, 
\cite{2020-Denoising-Autoencoders}, 
\cite{2020-Deinterleaving-Autoencoders}, 
\cite{2020-Deinterleaving-Automata}, this method does not require data to be digitized. There is no need to train a network for each target type and to iterate input and output data repeatedly. Very precise PRI values are not required as prior information.  This study also explored which network architecture is more advantageous among the existing NNs suitable for sequence modeling.
 
This paper is organized as follows. Section II introduces literature on radar signal deinterleaving and NNs for image semantic segmentation and sequence modeling. The characteristics and data model of this task are discussed in Section III. Based on the analysis presented in Sections II and III, we determine the NN architecture and deinterleaving strategy adopted by the proposed method in Section IV. Section V presents the experimental results and analysis. Section VI concludes the entire research.

\begin{figure}[!t]
\centering
\includegraphics[width=2.5in]{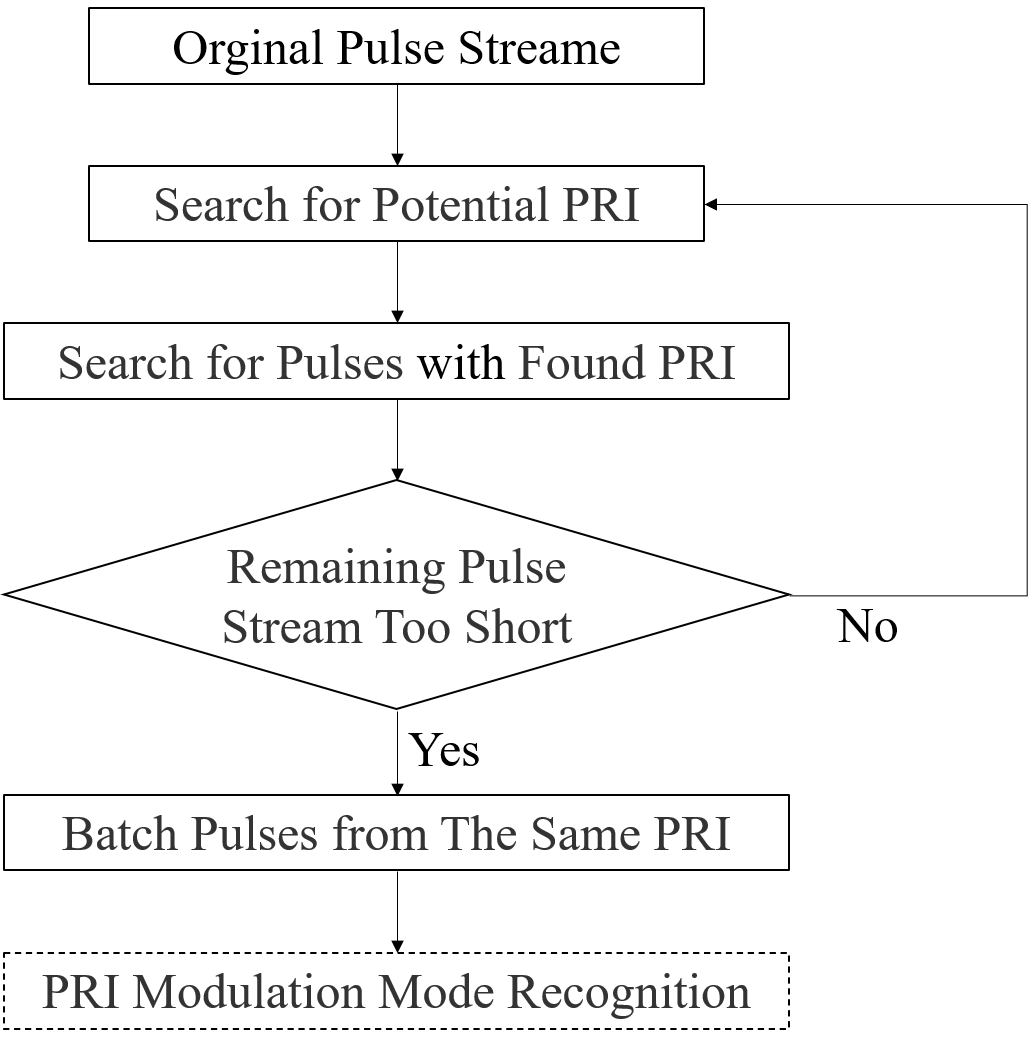}
\caption{Deinterleaving process of the traditional methods based on found PRI.}
\label{traditional methods}
\end{figure}

\begin{figure}[!t]
\centering
\includegraphics[width=1.8in]{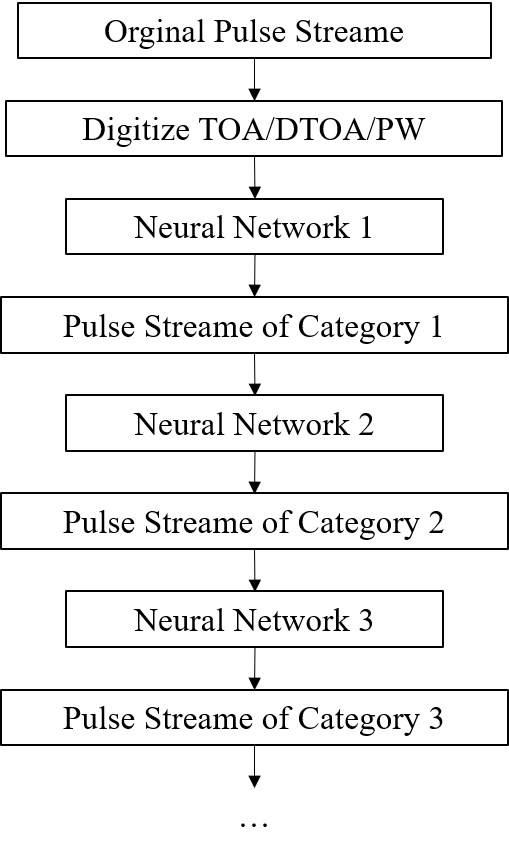}
\caption{Deinterleaving process of existing methods using neural nework.}
\label{neuralneworkmethods}
\end{figure}

\begin{figure}[!t]
\centering
\includegraphics[width=3.5in]{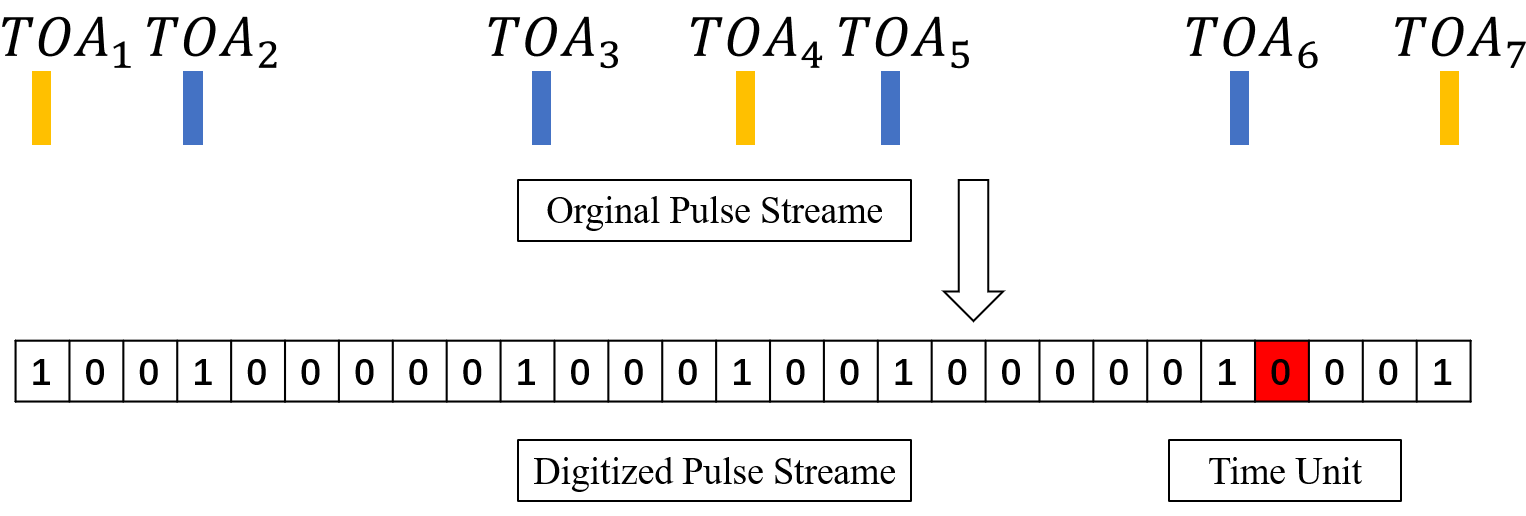}
\caption{Digitize TOA.}
\label{Digitize}
\end{figure}
%\begin{figure}[!t]
%\centering
%\includegraphics[width=2.5in]{Effectiveness of image semantic segmentation.png}
%\caption{Effectiveness of image semantic segmentation.}
%\label{Effectiveness}
%\end{figure}

\section{RELATED WORK}
\subsection{Radar signal deinterleaving}
The research of radar signal deinterleaving can be divided into two categories: deinterleaving based on multi-parameter and deinterleaving based on arrival time information. The former makes comprehensive use of the TOA, DOA, PW, RF, and PA,  while the latter uses only the TOA. This paper studies the deinterleaving method based on arrival time information. 

The radar PRI is the interval between fronts of adjacent pulses when the radar transmits signal. In the deinterleaving methods based on TOA, an important concept is to use the periodicity of radar PRI. This concept first finds the radar PRI or PRI period from information about the difference of TOA (DTOA) of the pulse stream and then uses the found PRI or PRI period to search the target radar pulse from the pulse stream, as shown in Fig. \ref{traditional methods}. 
\cite{1993-Wiley}, 
\cite{1989-CDIF}, 
\cite{1992-SDIF}, 
\cite{2000-PRITran}, 
\cite{1999-Spectrum-Estimation-of-Interleaved-Pulse-Trains}, 
\cite{2002-New-Matrix-Method}, 
\cite{1998-A-Novel-Pulse-TOA-Analysis-Technique}. 
Some methods use the histogram of DTOA to obtain radar PRI, such as cumulant difference histogram \cite{1989-CDIF} and sequential difference histogram (SDIF) 
\cite{1992-SDIF}, 
while others use DTOA matrix to find PRI 
\cite{2002-New-Matrix-Method}, 
\cite{1998-A-Novel-Pulse-TOA-Analysis-Technique}. Another method involves obtaining the spectrum of PRI through the transformation of arrival time and then extracting the real PRI 
\cite{2000-PRITran}, 
\cite{1999-Spectrum-Estimation-of-Interleaved-Pulse-Trains}. 
PRI transform algorithm has attracted much attention because of its excellent harmonic suppression performance (PRI-Tran) \cite{2000-PRITran}.

The above methods have several shortcomings. First, when radar signal pulses are dense or the target pulse loss rate is high, it is easy to find the wrong PRI. Second, when pulses belonging to different targets approach, it is difficult to distinguish them accurately. Third, when searching for potential PRI and target pulses, it is necessary to set thresholds and tolerances based on experience, making the deinterleaving effect prone to large fluctuations. Moreover, the threshold is very sensitive to the data quality and the deinterleaving effect may fluctuate significantly when the data quality changes. Fourth, these methods require iterating input and output data repeatedly to find new PRI and search pulses, which increases the complexity of the algorithm and requires more time. Especially, when the number of pulses in a period is large \cite{1989-CDIF}, 
\cite{1992-SDIF}, \cite{2000-PRITran}, the process of multiple rounds of searching for PRI period and pulses may consume a lot of time. Fifth, after the deinterleaving, the PRI modulation mode of the target radar signal is still unknown. So it is necessary to recogonize the modulation mode of target radar signal. 

In other methods, a radar pulse stream with periodical PRI is modeled as a linear dynamic model, and the Kalman filter is used for deinterleaving 
\cite{1994-Kalman}, 
\cite{1998-Kalman}, 
\cite{1999-Kalman}, 
\cite{2010-Kalman-MHT}. 
Some researchers have also attempped to apply the Hidden Markov Model to radar signal deinterleaving 
\cite{2005-HiddenMarkov}.
The algorithmic processes of these methods work only when some pre-assumptions are met for the streams, and they are complicated to be packaged for practical usages.

Researchers have attempted to use NNs to classify pulses very early, but since NNs were not fully developed at that time, they could only deal with relatively simple problems 
\cite{2007-Deinterleaving-of-radar-signals-and-PRF-identification-algorithms}, 
\cite{1991-Radar-signal-categorization-using-a-neural-network}, 
\cite{1998-self-organizing-networks-clustering-pulses}, 
\cite{2013-Radar-Emitter-Signals-Recognition-and-Classification-with-Feedforward-Networks}. 
Recurrent NNs (RNNs) have been introduced to deinterleave radar signals in 
\cite{2020-lzm-Classification-Denoising}. 
They treat the deinterleaving problem as a prediction problem, so only unidirectional information about the pulse sequence is used to judge the attribution of each pulse. Autoencoders are used to reduce noise pulses, but can not find the target pulses \cite{2020-Denoising-Autoencoders}. 
When autoencoders are used to find target pulses, accurate prior information about target pulse parameters is required 
\cite{2020-Deinterleaving-Autoencoders}. 
To facilitate NN processing of radar full pulse data, these methods 
\cite{2020-lzm-Classification-Denoising}, 
\cite{2020-Denoising-Autoencoders}, 
\cite{2020-Deinterleaving-Autoencoders} 
use small time units to digitize time information such as TOA, DTOA, PRI, and PW, as shown in Fig. \ref{Digitize}. This operation introduces three problems, which we call “resolution problems”: first, it introduces errors and reduces the accuracy of time information \cite{2020-lzm-Classification-Denoising}, 
\cite{2020-Denoising-Autoencoders}, 
\cite{2020-Deinterleaving-Autoencoders}; 
second, when there is more than one pulse in a time unit, only one pulse is presented, and information about other pulses is covered 
\cite{2020-lzm-Classification-Denoising}, 
\cite{2020-Denoising-Autoencoders}, 
\cite{2020-Deinterleaving-Autoencoders}. 
Third, the methods based on this operation may mark a position with no pulse as having one
\cite{2020-Denoising-Autoencoders}, 
\cite{2020-Deinterleaving-Autoencoders}. 
In addition, the above method needs to train a network for each category of radar signal, and only one target can be deinterleaved in each output step, as shown in Fig. \ref{neuralneworkmethods}. In other words, the existing method completes a binary classification task in each output step, so it is necessary to iterate input and output data repeatedly.

Finite automata have also been used for radar signal deinterleaving by researchers, but they also require prior information about target pulse parameters 
\cite{2020-Deinterleaving-Automata}.
\subsection{NNs for semantic segmentation and sequence modeling}
In recent years, the application of NNs in the field of image semantic segmentation has been thoroughly studied, and some important results have been produced 
\cite{2018-deeplabv3+}, 
\cite{2017-SegNet}, 
\cite{2015-deeplabv1}, 
\cite{2017-DeepLabv2}, 
\cite{2017-deeplabv3}, 
\cite{2015-FCN}, 
\cite{2017-RefineNet}, 
\cite{2015-U-Net}.
To achieve a good semantic segmentation effectiveness, some important concepts have been proposed, such as “fully convolutional network 
\cite{2015-FCN},” “multi-path refinement network,” “U-Net architecture 
\cite{2015-U-Net},” “encoder-decoder architecture 
\cite{2018-deeplabv3+}, 
\cite{2017-SegNet},” “fully connected conditional random field \cite{2015-deeplabv1},” and “atrous spatial pyramid pooling
\cite{2017-DeepLabv2}.” These concepts have achieved excellent effects in image semantic segmentation.

NNs are widely used in sequence modeling tasks. The RNN is the most popular sequence modeling architecture so far and has been considered the best architecture for a long time. People began studying the RNN model in the 1980s 
\cite{1990-Elman-Finding-Structure-in-Time}, 
\cite{1982-Hopfield-Neura-networks-and-physical-systems}, 
\cite{1986-Serial-order}, 
and proposing the Jordan Network in 1986 
\cite{1986-Serial-order}  and the Elman Network in 1990 
\cite{1990-Elman-Finding-Structure-in-Time}. The latter became the basis of some RNN architectures with a higher application value. In 1997, Jurgen Schmidhuber proposed the long short-term memory (LSTM) architecture 
\cite{1997-LSTM}, 
which uses gated unit and memory mechanism to improve RNNs in training. In the same year, Mike Schuster proposed a bidirectional RNN model (BRNN)
\cite{1997-BRNN}, which enables RNNs to simultaneously use sequence information in both forward and backward directions. The development of the gated recurrent unit (GRU) further improved the training problem of RNNs 
\cite{2014-GRU}. The application of RNN encoder—decoder effectively solves the sequence to sequence (seq2seq) problem 
\cite{2014-Seq2SequenceNN}, 
\cite{2014-Learning-Phrase-RNN-Encoder-Decoder-Translation}. The introduction of the attention-based models significantly improves the performance of the RNN-based model on many tasks 
\cite{2014-Neural-Machine-Translation-Jointly-Learning-Align-and-Translate}, 
\cite{2015-Attention-based-models-for-speech-recognition}. The use of transformer models pushes attention-based models to new heights while abandoning the RNN architecture 
\cite{2017-Attention}.

The application of convolutional NNs (CNNs) in sequence modeling tasks can also be traced back to the 1980s \cite{1987-Parallelnetworks}, 
\cite{1989-Connectionist-learning-procedures}, 
\cite{1990-PhonemeReco}. In recent years, CNN-based models have also performed excellently in some sequence modeling tasks, including audio synthesis, word-level language modeling, and machine translation, and can achieve state-of-the-art performance 
\cite{2016-Language-Modeling-Gated-Convolutional-Networks}, 
\cite{2017-convolutional-encoder-translation}, 
\cite{2017-Convolutional-sequence-to-sequence-learning}, 
\cite{2016-Neural-Machine-Translation-in-Linear-Time}, 
\cite{2016-WaveNet} in some tasks. These results prompt researchers to ponder: can the CNN architecture outperform the RNN architecture in more tasks, or is it simply limited to some specific tasks? Shaojie Bai et al. conducted an empirical study on this question \cite{2018-TCN-RNN} and proved that the CNN architecture outperformed the RNN architecture in many sequence modeling tasks, while these tasks are on the RNN’s “home Turf.” The authors summarize this CNN architecture as a temporal convolutional network (TCN) \cite{2018-TCN-RNN}.

\section{TASK CHARACTERISTICS AND DATA MODEL}	
\subsection{Input and output form}
For the input data to have smaller variance and facilitate NN processing, the proposed method inputs DTOA of the pulse stream into the NN instead of TOA. When the pulse stream contains only a single radar target, the DTOA of the pulse stream is the real PRI of the target. When the pulse stream contains pulses from multiple radiation sources or pulses from a single target arrive through multiple paths, the DTOA of the pulse stream is chaotic. To make DTOA and TOA equal in length, we add 0 before DTOA as the first value of DTOA. The output of the proposed method is the label information about the category of each pulse. When training the NN, the DTOA and pulse labels to input are into the network, as shown in Fig. \ref{interleaved}.

Unlike existing deinterleaving methods based on NNs 
\cite{2020-lzm-Classification-Denoising}, 
\cite{2020-Denoising-Autoencoders}, 
\cite{2020-Deinterleaving-Autoencoders}, the proposed method does not digitize time information as input to avoid adverse effects caused by it. In the output, each pulse is judged to determine the target category to which it belongs, rather than whether it belongs to the target we want. In other words, the processing result of the proposed method is multi-classification, whereas that of existing methods is typically binary classification.
\subsection{Difference between this task and image semantic segmentation and other sequence modeling tasks}
Radar signal deinterleaving based on semantic segmentation is a problem of mapping an input sequence to an output sequence. It is different from image semantic segmentation and seq2seq tasks such as natural language processing (NLP).
\subsubsection{Target points are unconcentrated and throughout the sequence}
In the pulse stream, the pulses of different targets are interleaved and information about the same target runs through the entire pulse stream. However, in image semantic segmentation, the pixels of the same object are typically concentrated in one or several regions.
\subsubsection{There is a strict mathematical relationship between the data at each input point of the sequence}
Since the input is DTOA, the information loss of one data point completely changes the information in the pulse stream, so pooling is not allowed. Image semantic segmentation and seq2seq tasks do not have this feature.
\subsubsection{The input and output are of equal length}
The input and output of this task are equal length sequences. In some sequence modeling tasks such as machine translation, the input and output are often not of equal length.
\subsubsection{The input at each point is meaningless on its own}
Similar to image data, the input data of each point in this task are meaningless alone. Only when they are computed along with other input data, can their information be reflected. However, in NLP tasks, each input word has a specific meaning.
\subsubsection{Forward and backward information is equivalent}
In the deinterleaving task, the forward and reverse information about a sequence is completely equivalent, which is significantly different from many sequence modeling tasks. Therefore, in this task, using bidirectional information about the sequence simultaneously is more conducive to accurately judging the category of each pulse.

\subsection{Limitations of the deinterleaving method based on semantic segmentation and the solution}
\begin{figure}[!t]
\centering
\includegraphics[width=2.5in]{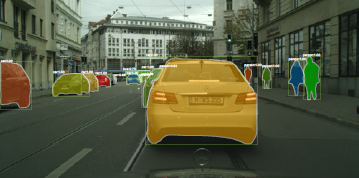}
\caption{Effectiveness of image instance segmentation.}
\label{mask_r-cnn}
\end{figure}

The deinterleaving method based on semantic segmentation faces the same problem as image semantic segmentation, that is, it cannot distinguish multiple objects belonging to the same category in a single input. In image processing, this problem is solved by instance segmentation 
\cite{2017-Mask-R-CNN}, 
\cite{2018-Path-Aggregation-Network-for-Instance-Segmentation}, 
\cite{2020-SOLO}, which typically consists of two contents: semantic segmentation and object detection, as shown in Fig. \ref{mask_r-cnn} \cite{2017-Mask-R-CNN}. 
However, this solution cannot be applied to radar signal deinterleaving because, in an image, pixels belonging to the same target are concentrated, whereas in a pulse stream, pulses belonging to the same target are not. Pulses from one target are interleaved with pulses from other targets and distributed in the entire pulse stream. The method to solve this problem is to extract more different semantics from target radar signals and divide radar signals into more classes. This will be discussed in section IV.

\subsection{Data model}
\begin{figure}[!t]
\centering
\subfloat[]{\includegraphics[width=1.6in]{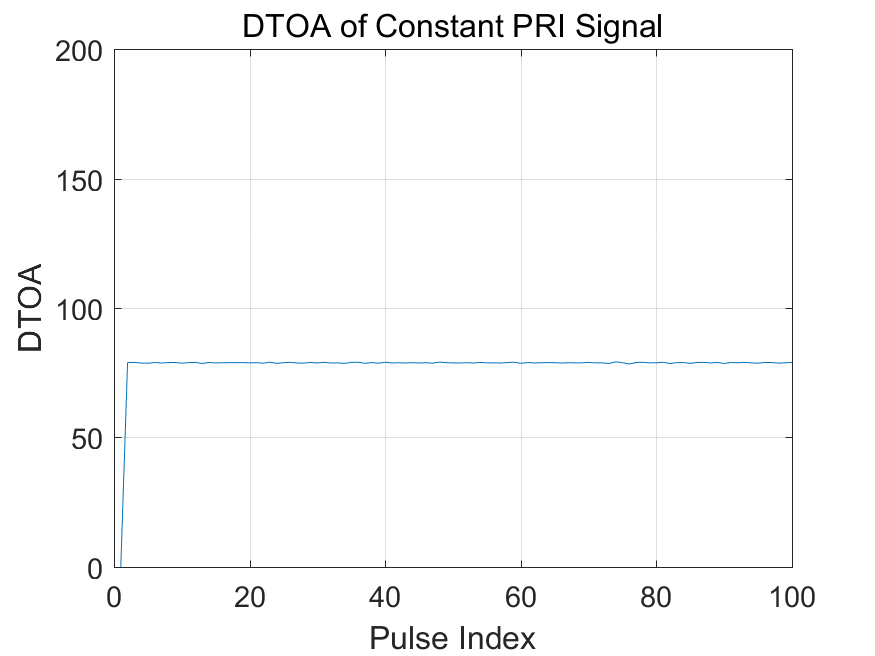}%
\label{ConstantPRI-a}}
\hfil
\subfloat[]{\includegraphics[width=1.6in]{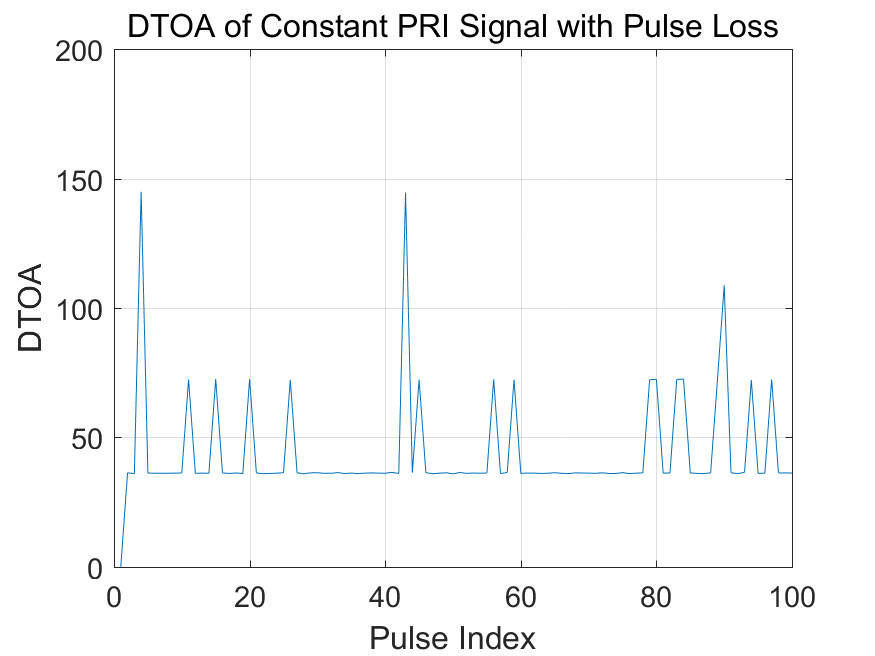}%
\label{ConstantPRI-b}}
\hfil
\subfloat[]{\includegraphics[width=1.6in]{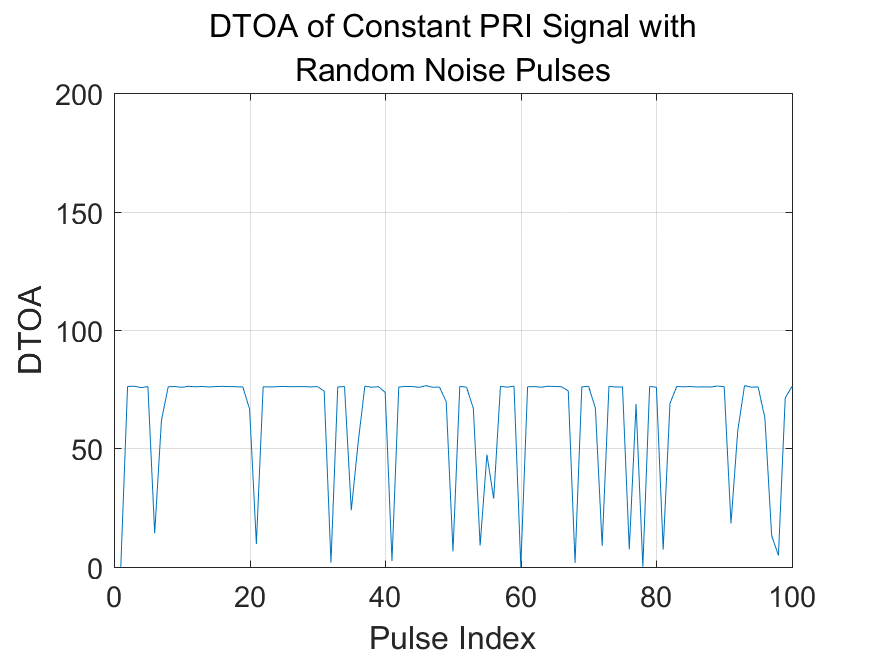}%
\label{ConstantPRI-c}}
\hfil
\subfloat[]{\includegraphics[width=1.6in]{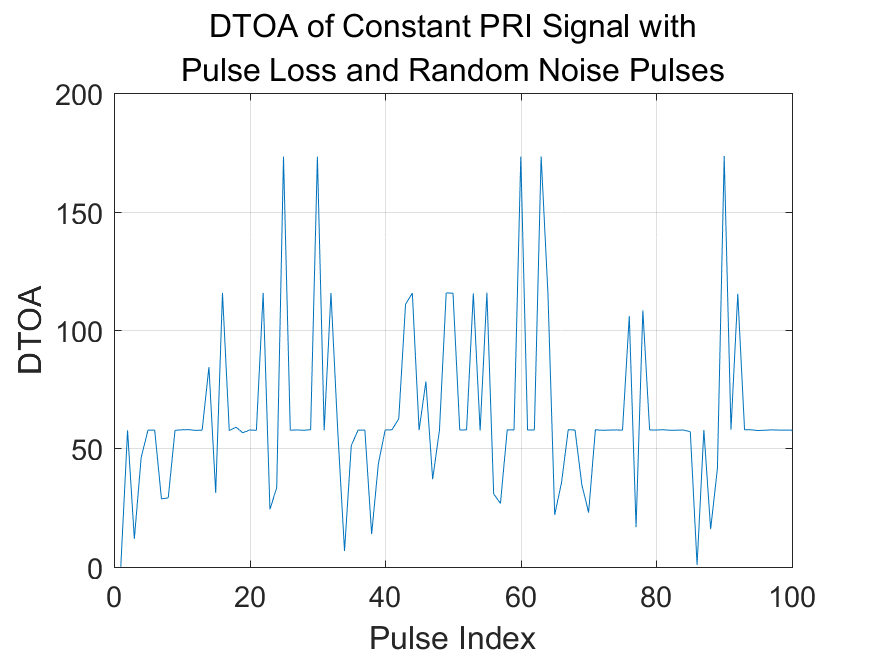}%
\label{ConstantPRI-d}}
\caption{DTOA of Radar Signal with Constant PRI}
\label{ConstantPRI}
\end{figure}

\begin{figure}[!t]
\centering
\subfloat[]{\includegraphics[width=1.6in]{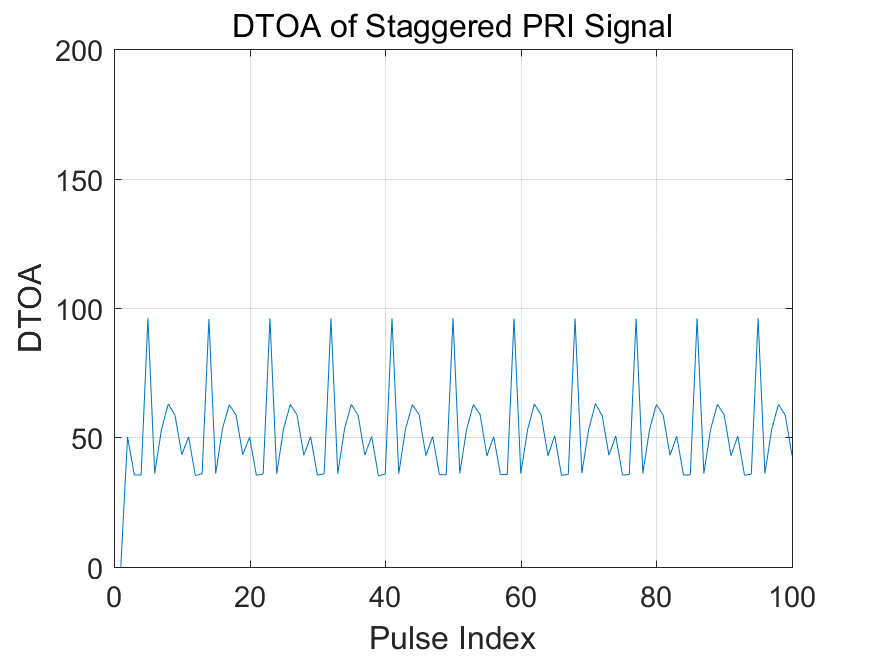}%
\label{StaggeredPRI-a}}
\hfil
\subfloat[]{\includegraphics[width=1.6in]{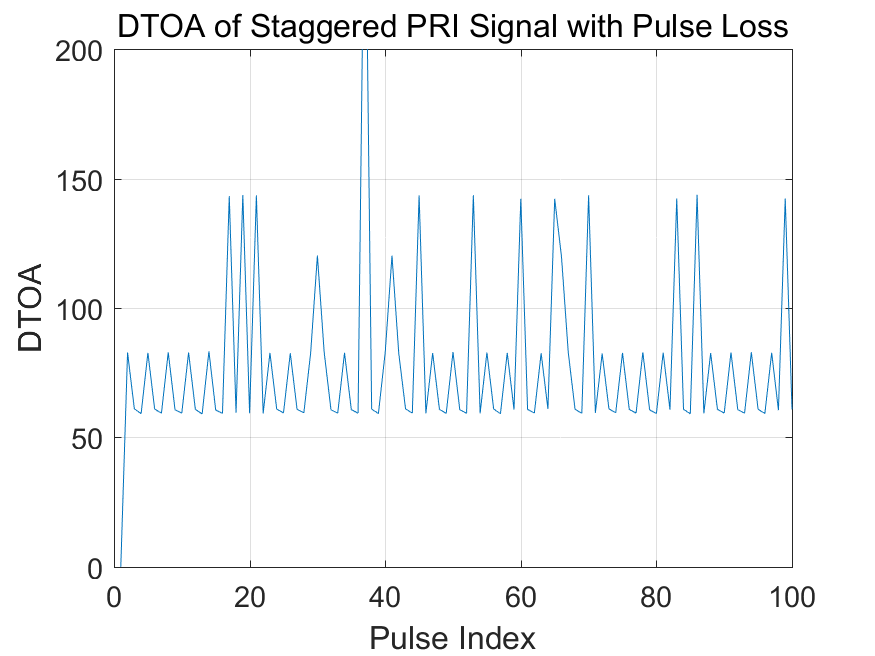}%
\label{StaggeredPRI-b}}
\hfil
\subfloat[]{\includegraphics[width=1.6in]{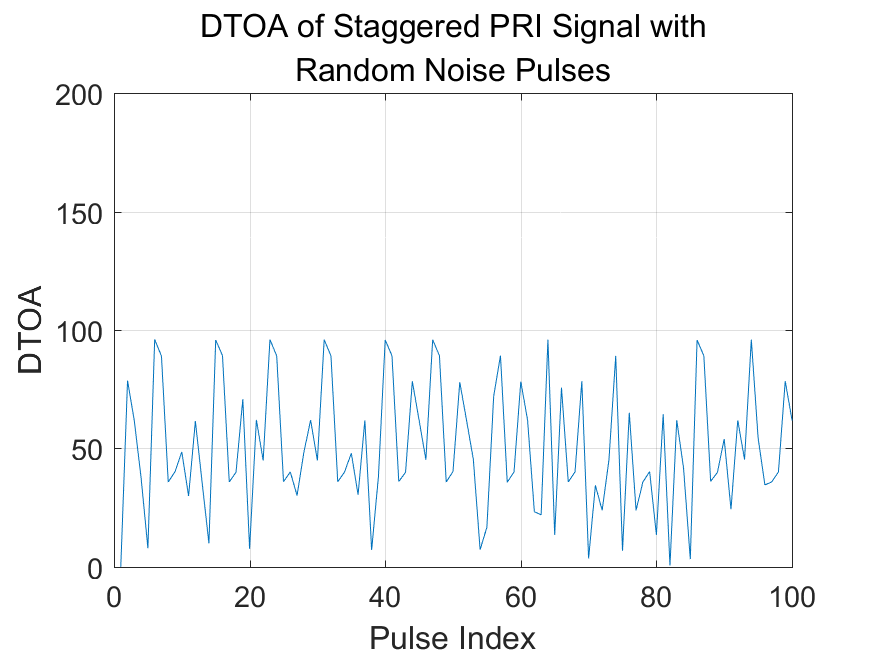}%
\label{StaggeredPRI-c}}
\hfil
\subfloat[]{\includegraphics[width=1.6in]{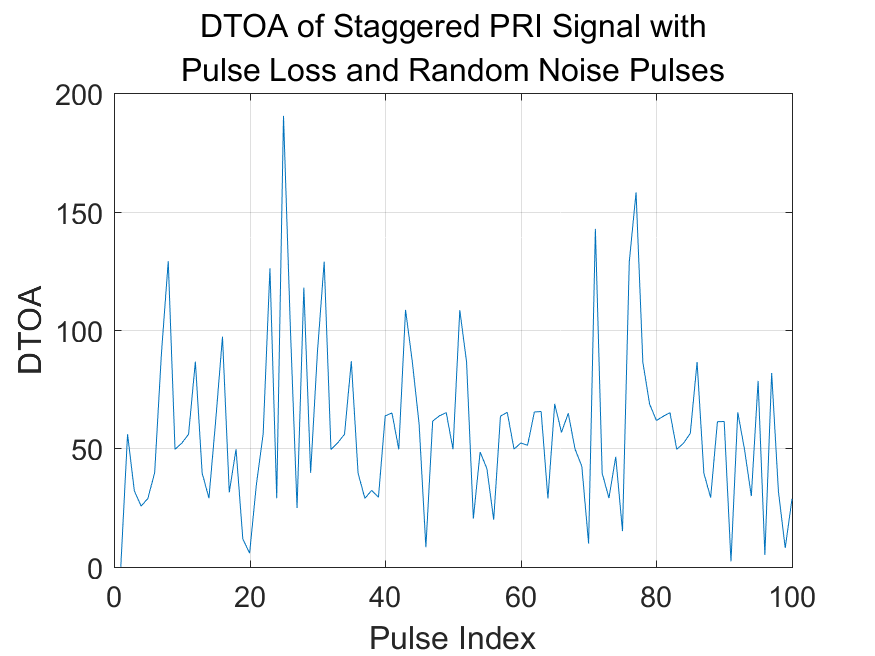}%
\label{StaggeredPRI-d}}
\caption{DTOA of Radar Signal with Staggered PRI}
\label{StaggeredPRI}
\end{figure}

PRI modulation mode has an important effect on radar function and performance. In this paper, we define and use three PRI modulation modes in simulation experiments. The DTOA of radar signals with these three PRI modulation modes is shown in Figs. \ref{ConstantPRI}—\ref{StaggeredPRI}. The subfigures show the DTOA of radar signals under different conditions: a) DTOA of nondestructive radar signal, that is, PRI of radar signal; b) DTOA of radar signal with pulse loss; c) DTOA of radar signal with random noise pulses; and d) DTOA of radar signal with pulse loss and random noise pulses.
\subsubsection{Constant PRI}
The radar PRI remains a constant, and the PRI sequence can be represented as
%$$ {{PRI}_{n}}={{PRI}_{0}},n=1,2,3...   \tag{1}$$
\begin{equation}
{{PRI}_{n}}={{PRI}_{0}},n=1,2,3.... \label{con for}
\end{equation}

\subsubsection{Dwell and Switch (D\&S) PRI}
Radar PRI changes in groups, with the same number of pulses in each group. The value of PRI changes periodically between groups. Its mathematical model is expressed as
\begin{equation}
PR{{I}_{n}}=PR{{I}_{n+j}},0\le j<J, \label{D&S for1}
\end{equation}

\begin{equation}
{{PRI}_{n}}={{PRI}_{n+N*K}}. \label{D&S for2}
\end{equation}
${{PRI}_{n}} $ is the first PRI in each group, ${J}$ represents the number of pulses in each group, and ${K}$ represents the number of pulse groups in one period, that is, the number of PRI values in a period.

\subsubsection{Staggered PRI}
Radar PRI consists of several fixed values and changes periodically. The PRI sequence can be described as follows:
\begin{equation}
PR{{I}_{n}}=PR{{I}_{n+M}}. \label{sta for}
\end{equation}
${M}$ represents the number of PRI values in a period.

\begin{figure}[!t]
\centering
\subfloat[]{\includegraphics[width=1.6in]{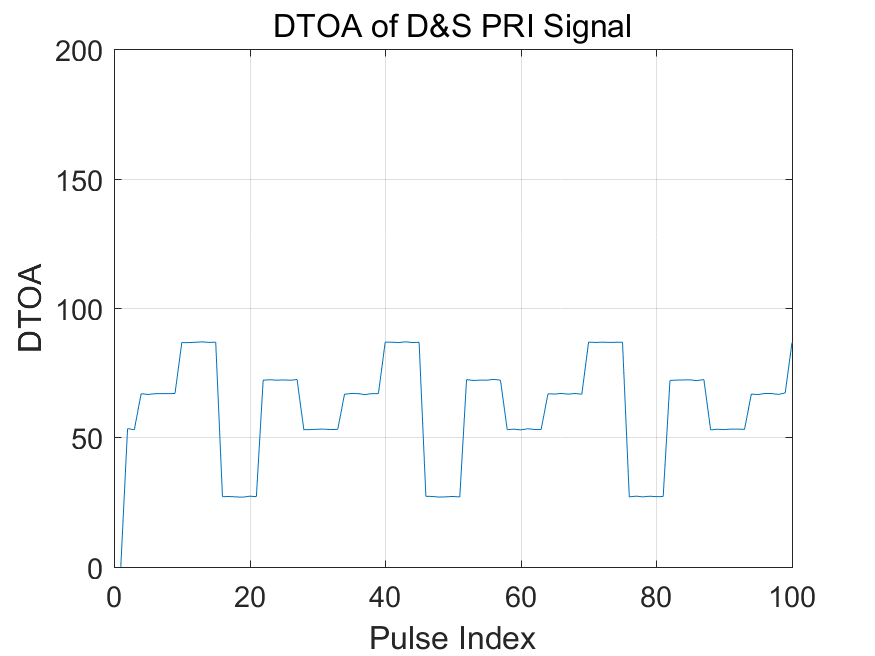}%
\label{D_S-a}}
\hfil
\subfloat[]{\includegraphics[width=1.6in]{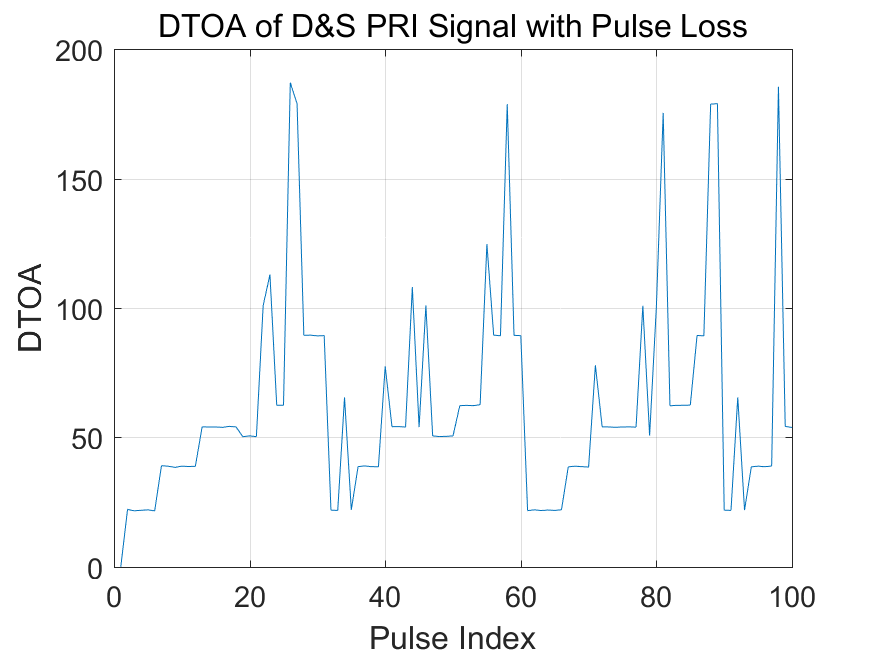}%
\label{D_S-b}}
\hfil
\subfloat[]{\includegraphics[width=1.6in]{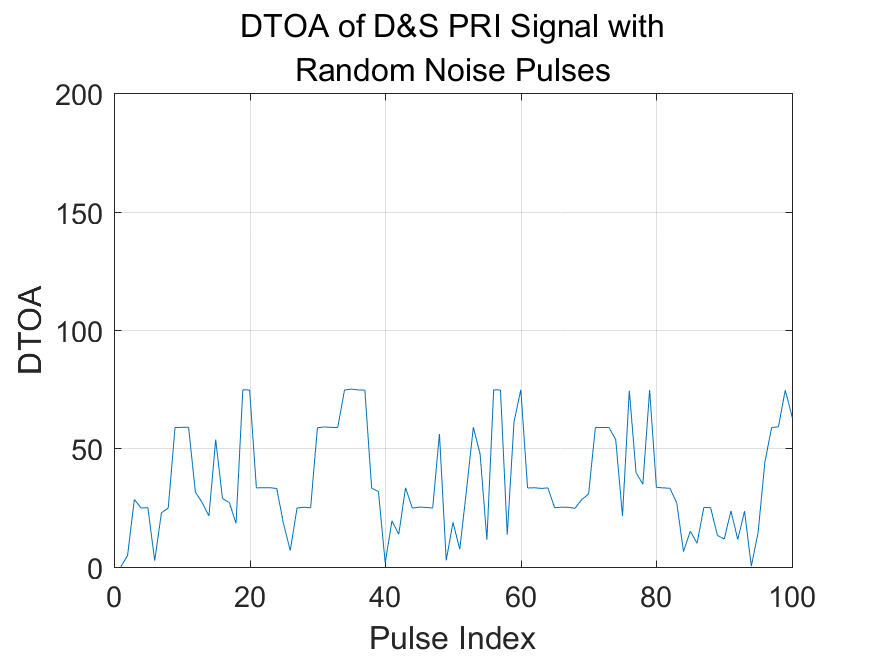}%
\label{D_S-c}}
\hfil
\subfloat[]{\includegraphics[width=1.6in]{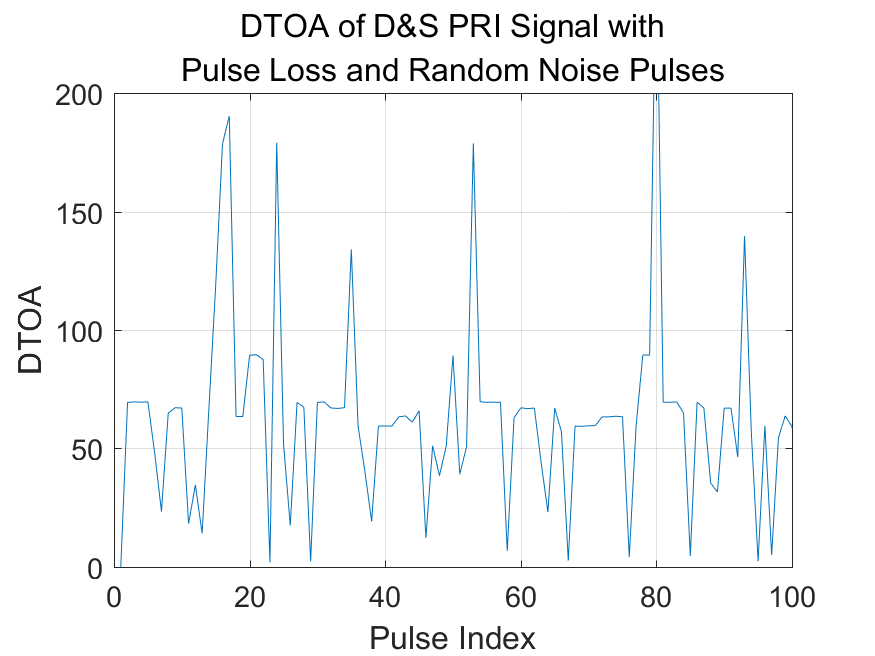}%
\label{D_S-d}}
\caption{DTOA of Radar Signal with D\&S PRI}
\label{D_S}
\end{figure}

\section{NEURAL NETWORK ARCHITECTURE AND DEINTERLEAVING STRATEGY}

\subsection{What kind of NN should be selected for this task}
According to the concept of the proposed method, we select a NN, and it is required to have good semantic segmentation ability for sequence data with strong mathematical relations. In Section III, we analyzed the difference between this task and image semantic segmentation and seq2seq tasks such as NLP. 

Therefore, the NN used in the proposed method needs to meet the following requirements: good ability at sequence modeling, full use of all information about the entire sequence (or a sufficiently large receptive field), equal length of input and output, and no pooling.

Accordingly, we select the BRNN and dilated convolutional network (DCN) but ignore the classical NNs used for image semantic segmentation 
\cite{2018-deeplabv3+}, 
\cite{2017-SegNet}, 
\cite{2015-deeplabv1}, 
\cite{2017-DeepLabv2}, 
\cite{2017-deeplabv3}, 
\cite{2015-FCN}, 
\cite{2017-RefineNet}, 
\cite{2015-U-Net} and used for seq2seq tasks, 
\cite{2014-Seq2SequenceNN}, 
\cite{2014-Learning-Phrase-RNN-Encoder-Decoder-Translation}, 
\cite{2014-Neural-Machine-Translation-Jointly-Learning-Align-and-Translate}, 
\cite{2015-Attention-based-models-for-speech-recognition}, 
\cite{2017-Attention}, e. g. encoder—decoder structures architecture.
\subsubsection{BRNN}
\begin{figure}[!t]
\centering
\includegraphics[width=3in]{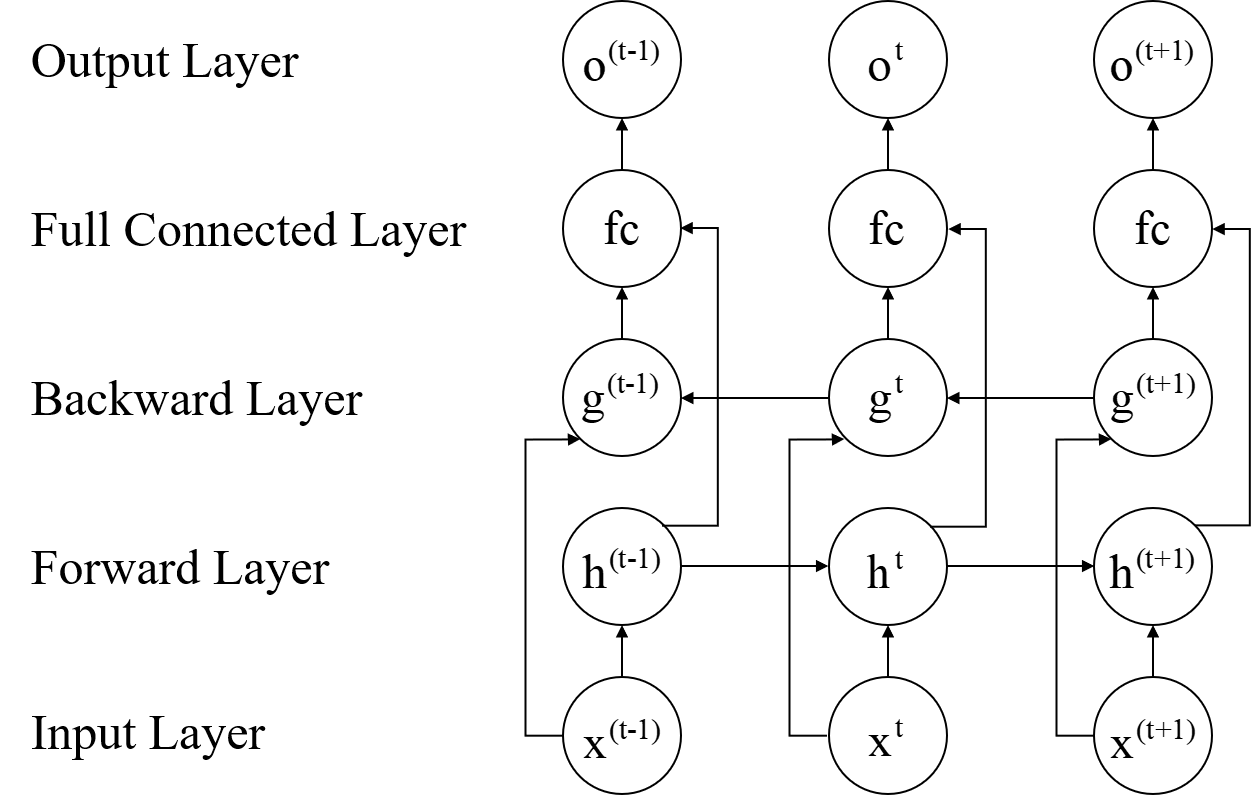}
\caption{Deinterleaving process of BRNN.}
\label{BRNN}
\end{figure}

For DTOA data, forward and reverse information is equivalent. To make full use of the complete information about the sequence when determining the category of each pulse, the BRNN is used in this study to process DTOA data. Then, the output of each step of the BRNN is connected with the full connected layer to realize the classification of each time step, as shown in Fig. \ref{BRNN}. This study uses the LSTM and GRU architectures of RNN, that is, bidirectional GRU (BGRU) and LSTM (BLSTM), to achieve this task.
\subsubsection{DCN}
\begin{figure}[!t]
\centering
\includegraphics[width=3in]{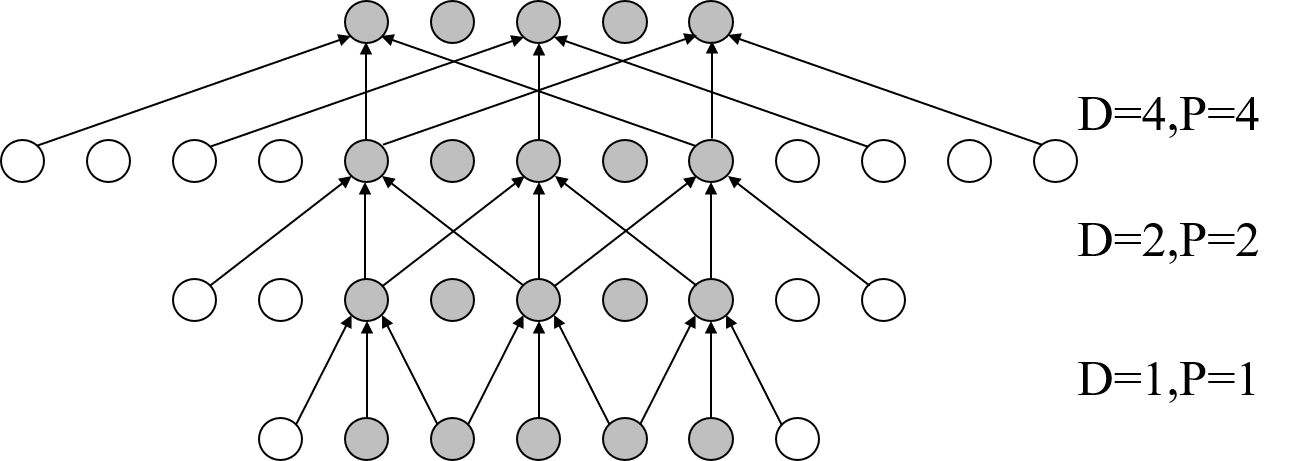}
\caption{Dilation and padding of DCN.}
\label{Dilation}
\end{figure}

Owing to TCN’s 
\cite{2018-TCN-RNN} 
outstanding performance in sequence modeling tasks, this study used TCN as a reference when constructing a DCN. In terms of residual module construction, a DCN is generally the same as a TCN. The only difference is that, to keep the length of the feature map unchanged in each convolution step, both sides of the feature map are padded symmetrically rather than adopting causal convolution, as shown in Fig. \ref{Dilation}. This operation considers the equivalence of forward and reverse information. In this study, we set the convolution kernel size of the DCN as 3 and 8 residual modules. This makes the receptive field of each convolution kernel in the last layer of the network sufficient to cover the length of the input data. After the residual module, we use a common convolution to reduce the number of channels to the number of target classes to achieve the classification of each pulse.
%\begin{figure}[!t]
%\centering
%\includegraphics[width=2.5in]{Residual block of DCN.png}
%\caption{Residual block of DCN.}
%\label{Residual}
%\end{figure}

\subsection{How to use semantic information—deinterleaving strategy}
Semantic segmentation is based on the characteristics of different categories of objects. Analysis in Section III indicates that sufficient different semantics need to be extracted from radar signals to achieve a good deinterleaving effect. This part mainly analyzes how to deinterleave radar signals using PRI modulation modes and PRI parameters as semantics separately and how to use both for deinterleaving comprehensively.
\subsubsection{Take PRI modulation modes as semantic information}
Different PRI modulation modes represent different information, that is, different semantics, as shown in Table \ref{Semantic}. The target radar signals can be divided into different categories accordingly. When the intercepted radar pulse stream contains multiple targets with different PRI modulation modes, PRI modulation information can be used as semantic information, and the category of each pulse can be predicted on the basis of this to achieve radar signal deinterleaving.
\subsubsection{Take PRI parameters as semantic information}
When multiple target radars adopt the same PRI modulation mode, the deinterleaving method based on semantic segmentation with PRI modulation modes is limited and cannot distinguish such multiple targets. In this case, PRI parameter information can be used as semantics to distinguish different targets. That is, the radar signal with PRI value located in (a, b) is the first subclass, the radar signal located in (b, c) is the second subclass, the radar signal located in (c, d) is the third subclass, and so on, as shown in Table \ref{Semantic}. How to set the specific value range of the subclass depends on the signal and the specific task environment.

\begin{table}[]
\centering
\caption{Semantic information for radar signal deinterleaving}
\label{Semantic}
\resizebox{3in}{!}{
\begin{tabular}{|c|l|}
\hline
\textbf{semantic information} & \multicolumn{1}{c|}{\textbf{category}}                                                                                        \\ \hline
PRI modulation mode           & \begin{tabular}[c]{@{}l@{}}category 1: constant PRI   \\ category 2: D\&S PRI\\ category 3: staggered PRI   \\ …\end{tabular} \\ \hline
PRI value                     & \begin{tabular}[c]{@{}l@{}}category 1: (a, b)\\ category 2: (b, c)\\ category 3: (c, d)\\ …\end{tabular}                         \\ \hline
\end{tabular}
}
\end{table}

\subsubsection{Comprehensive use of PRI modulation modes and PRI parameters}
\begin{figure*}[!t]
\centering
\includegraphics[width=6in]{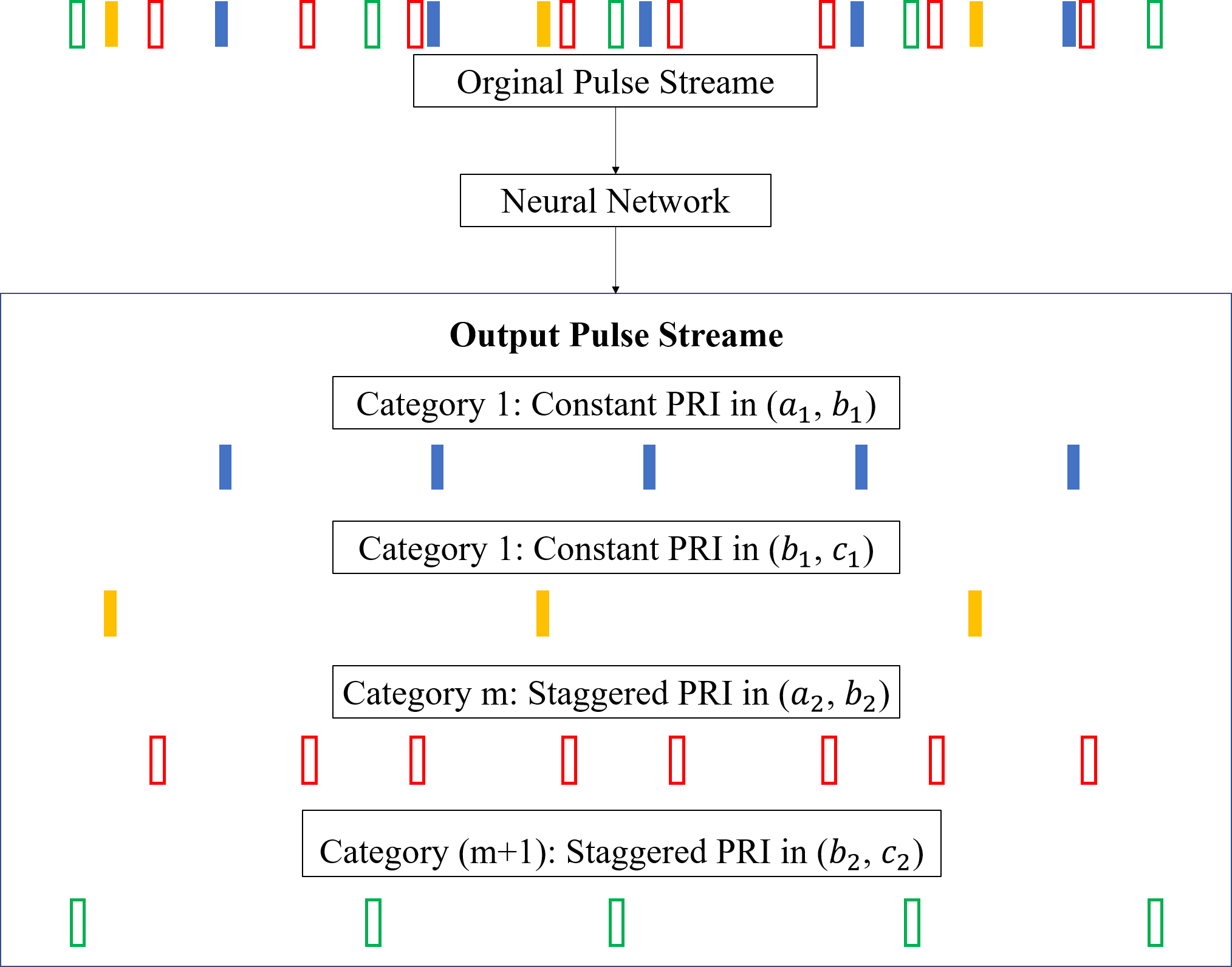}
\caption{Parallel deinterleaving method.}
\label{Parallel}
\end{figure*}

\begin{figure*}[!t]
\centering
\includegraphics[width=6in]{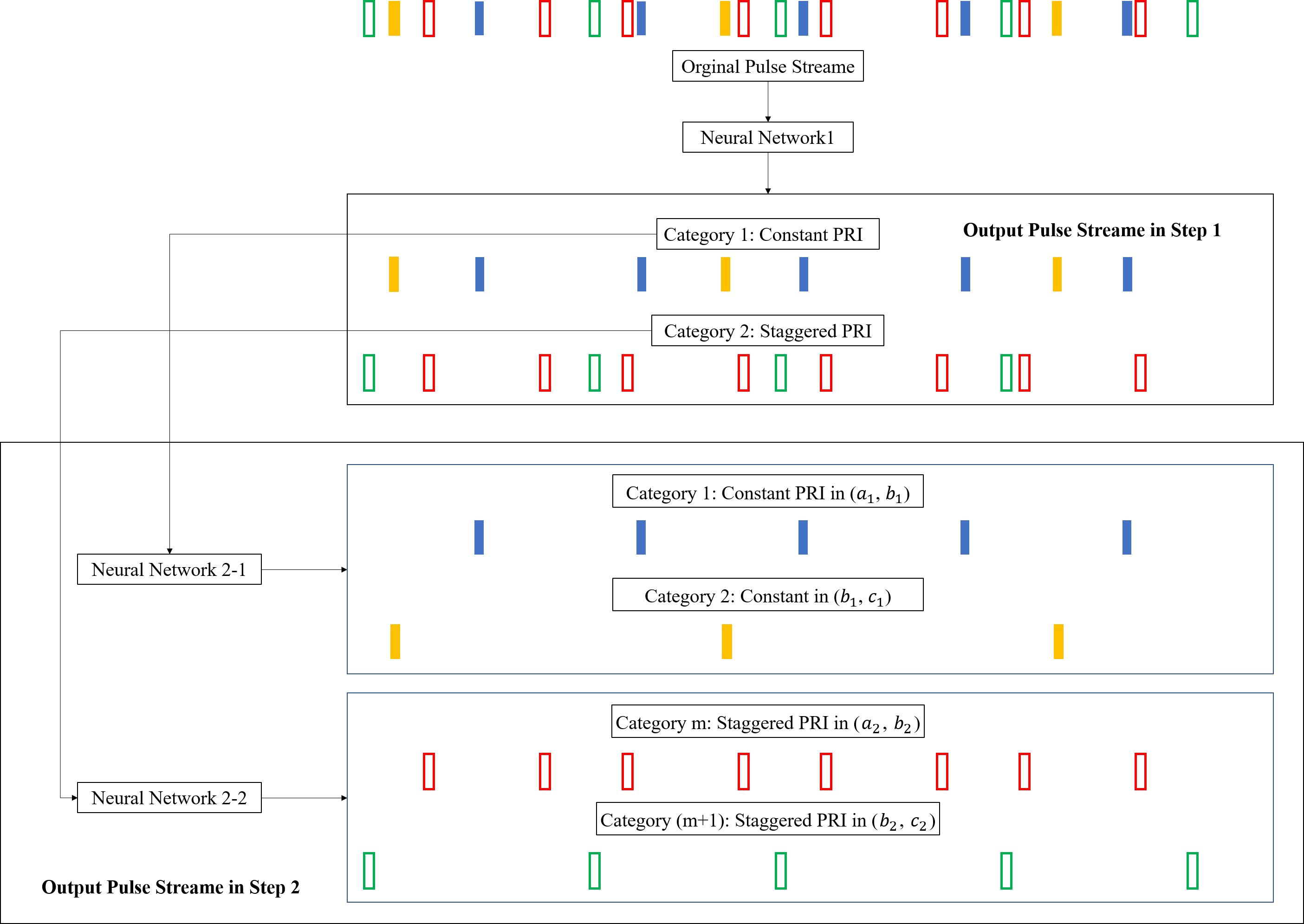}
\caption{Serial deinterleaving method.}
\label{Serial}
\end{figure*}

It was highlighted in Part C of Section III that we need sufficient semantics to divide radar pulses into different classes to solve the problem of having multiple targets in the same class. Here, we have proposed deinterleaving methods using PRI modulation and PRI parameters as semantic information. In this section, two methods are proposed to make comprehensive use of this information for deinterleaving. One is a parallel deinterleaving method, using PRI modulation modes and PRI parameters simultaneously, as shown in Fig. \ref{Parallel}. The second is the serial deinterleaving method, which first uses PRI modulation and then uses PRI parameters for deinterleaving, as shown in Fig. \ref{Serial}. The former has fewer deinterleaving steps and needs only one NN. The latter needs to be completed step by step and uses multiple NNs. However, the latter can adapt to more complex deinterleaving environments, e. g., when the semantic categories of radar pulses are diverse and the capacity of NNs is limited.

\subsection{Loss function}
In this task, each sample input to the NN is the DTOA of a pulse stream. The predicted loss of each sample by the NN is the average of the predicted loss of all pulses in the pulse stream, i.e.,

\begin{equation}
Loss=\frac{1}{N}\sum\nolimits_{n=1}^{N}{los{{s}_{n}}}. \label{LOSS}
\end{equation}
$los{{s}_{n}}$ is the predicted loss of the nth pulse by the NN. We use a cross-entropy loss function to evaluate the prediction performance of each pulse of the NN, which can be described below:

\begin{equation}
loss=-\sum\nolimits_{c=1}^{C}{{{P}_{c}}\log (\overset{\wedge }{\mathop{{{P}_{c}}}}\,)}. \label{loss}
\end{equation}
$C$ represents the category number of radar signals in the pulse stream. ${{P}_{c}}$ denotes whether the current pulse belongs to the cth radar signal category, and its value is either 0 or 1. $\overset{\wedge }{\mathop{{{P}_{c}}}}\,$ represents the probability that the current pulse belongs to the cth radar signal category in the NN’s prediction.
\section{EXPERIMENTS}

\subsection{Data simulation}
According to the definition of PRI modulation modes in Part D of Section III, the following designs are developed for simulation data.

1) For all PRI modulation modes, the PRI value satisfies the condition $20<PRI<100$ unless we specify it.

2) For the D\&S PRI, the number of pulses in each group satisfies the condition $4\le J\le 6$, and the number of pulse groups in one period satisfies the condition $4\le K\le 6$.

3) For the staggered PRI, the number of PRI values in a period satisfies the condition $3\le M\le 10$ unless we specify it.

4) In this paper, the Gaussian distributed deviation is added to TOA to simulate measurement errors, and the standard deviation is 0.1. Then, the DTOA is generated on this basis, and the length of the DTOA is 1,000.

5) In this paper, the problem of target pulse loss and random noise pulses in intercepted pulse stream is considered. The pulse loss rate of the target is represented by ${{\rho }_{l}}$, and the ratio of the number of random noise pulses to the average number of the target radars pulses in the intercepted pulse stream is represented by ${{\rho }_{n}}$. The proportion of the number of random noise pulses to the total number of pulses can be calculated by $\frac{{{\rho }_{n}}}{{{\rho }_{n}}+D}$, and $D$ represents the number of target radars.

\subsection{Design of experiments}
According to the deinterleaving strategy proposed in Section IV, five experiments are designed to verify the feasibility of the proposed deinterleaving method and compare the performance of different NNs in addressing this problem.

Experiment 1 verifies the feasibility of deinterleaving radar signal taking PRI modulation modes as semantics. Experiments 2 and 3 verify the feasibility of deinterleaving radar signal taking PRI parameters as semantics. Experiment 4 verifies the parallel deinterleaving method using PRI modulation modes and PRI parameters simultaneously. Experiment 5 is used to verify the first step of the serial deinterleaving method, deinterleaving radar signal with PRI modulation modes when there are multiple targets per PRI modulation mode. Experiments 2 and 3 verify the second step. ${{\rho }_{l}}$ and ${{\rho }_{n}}$ of each sample of the training data are randomly chosen within a certain range.
\subsubsection{Experiment 1—Deinterleaving radar signal with PRI modulation modes}
The target settings are shown in Table \ref{exp1tar}. For the training data, ${0<{\rho }_{l}<0.25}$, ${0<{\rho }_{n}<0.25}$.
\begin{table}[]
\centering
\caption{Target settings in Experiment 1}
\label{exp1tar}
\resizebox{3in}{!}{
\begin{tabular}{|c|c|c|}
\hline
\textbf{category} & \textbf{PRI modulation mode} & \textbf{number of targets} \\ \hline
1                 & constant                     & 1                          \\ \hline
2                 & D\&S                         & 1                          \\ \hline
3                 & staggered                    & 1                          \\ \hline
4                 & \multicolumn{2}{c|}{random noise pulse}                   \\ \hline
\end{tabular}
}
\end{table}

\subsubsection{Experiment 2—Deinterleaving radar signal of constant PRI with PRI values}
The target settings are shown in Table \ref{exp2tar}. For the training data, ${0<{\rho }_{l}<0.5}$, ${0<{\rho }_{n}<0.5}$.

\begin{table}[]
\centering
\caption{Target settings in Experiment 2}
\label{exp2tar}
\resizebox{3in}{!}{
\begin{tabular}{|c|cc|}
\hline
\textbf{category} & \multicolumn{1}{c|}{\textbf{\begin{tabular}[c]{@{}c@{}}value range of\\ constant PRI\end{tabular}}} & \textbf{number of targets} \\ \hline
1 & \multicolumn{1}{c|}{(20,40)}     & 1    \\ \hline
2 & \multicolumn{1}{c|}{(40,60)}     & 1    \\ \hline
3 & \multicolumn{1}{c|}{(60,80)}     & 1    \\ \hline
4 & \multicolumn{1}{c|}{(80,100)}    & 1    \\ \hline
5 & \multicolumn{2}{c|}{random noise pulse} \\ \hline
\end{tabular}
}
\end{table}

\subsubsection{Experiment 3—Deinterleaving radar signal of staggered PRI with PRI values}
The target settings are shown in Table \ref{exp3tar}. For the training data, ${0<{\rho }_{l}<0.5}$, ${0<{\rho }_{n}<0.5}$.

\begin{table}[]
\centering
\caption{Target settings in Experiment 3}
\label{exp3tar}
\resizebox{3in}{!}{
\begin{tabular}{|c|ccc|}
\hline
\textbf{category} &
  \multicolumn{1}{c|}{\textbf{\begin{tabular}[c]{@{}c@{}}value range of\\ staggered PRI\end{tabular}}} &
  \multicolumn{1}{c|}{\textbf{\begin{tabular}[c]{@{}c@{}}number of pulses\\ in a period\end{tabular}}} &
  \textbf{number of targets} \\ \hline
1 & \multicolumn{1}{c|}{(20,40)}  & \multicolumn{1}{c|}{7} & 1 \\ \hline
2 & \multicolumn{1}{c|}{(40,70)}  & \multicolumn{1}{c|}{7} & 1 \\ \hline
3 & \multicolumn{1}{c|}{(70,100)} & \multicolumn{1}{c|}{7} & 1 \\ \hline
4 & \multicolumn{3}{c|}{random noise pulse}                    \\ \hline
\end{tabular}
}
\end{table}

\subsubsection{Experiment 4—Deinterleaving radar signal using PRI modulation modes and PRI parameters simultaneously}
The target settings are shown in Table \ref{exp4tar}. For the training data, ${0<{\rho }_{l}<0.5}$, ${0<{\rho }_{n}<0.5}$.

\begin{table}[]
\centering
\caption{Target settings in Experiment 4}
\label{exp4tar}
\resizebox{3in}{!}{
\begin{tabular}{|c|c|c|}
\hline
\textbf{category} & \textbf{\begin{tabular}[c]{@{}c@{}}PRI modulation mode \\ and value range\end{tabular}} & \multicolumn{1}{c|}{\textbf{number of targets}} \\ \hline
1 & constant PRI in (20,60)       & 1       \\ \hline
2 & constant PRI in (60,100)       & 1       \\ \hline
3 & staggered PRI                 & 1       \\ \hline
4 & \multicolumn{2}{c|}{random noise pulse} \\ \hline
\end{tabular}
}
\end{table}

\subsubsection{Experiment 5—Deinterleaving radar signal with PRI modulation modes when there are multiple targets per PRI modulation mode}
The target settings are shown in Table \ref{exp5tar}. For the training data, ${0<{\rho }_{l}<0.25}$, ${0<{\rho }_{n}<0.25}$.

\begin{table}[]
\centering
\caption{Target settings in Experiment 5}
\label{exp5tar}
\resizebox{3in}{!}{
\begin{tabular}{|c|ccc|}
\hline
\textbf{category} &
  \multicolumn{1}{c|}{\textbf{\begin{tabular}[c]{@{}c@{}}PRI modulation mode   \\ and value range\end{tabular}}} &
  \multicolumn{1}{c|}{\textbf{\begin{tabular}[c]{@{}c@{}}number of pulses\\ in a period\end{tabular}}} &
  \textbf{\begin{tabular}[c]{@{}c@{}}number of\\ targets\end{tabular}} \\ \hline
\multirow{2}{*}{1} & \multicolumn{1}{c|}{constant PRI in (20,60)}   & \multicolumn{1}{c|}{1} & 1 \\ \cline{2-4} 
                   & \multicolumn{1}{c|}{constant PRI in (60,100)}  & \multicolumn{1}{c|}{1} & 1 \\ \hline
\multirow{2}{*}{2} & \multicolumn{1}{c|}{staggered PRI in (20,60)}  & \multicolumn{1}{c|}{7} & 1 \\ \cline{2-4} 
                   & \multicolumn{1}{c|}{staggered PRI in (60,100)} & \multicolumn{1}{c|}{7} & 1 \\ \hline
3                  & \multicolumn{3}{c|}{random noise pulse}                                     \\ \hline
\end{tabular}
}
\end{table}

\subsection{Result}
The network capacity in the five experiments and the overall performance on the test set (produced under the same conditions as the training set) are listed in Table \ref{capacity}.

\begin{figure*}[!t]
\centering
\subfloat[]{\includegraphics[width=2.3in]{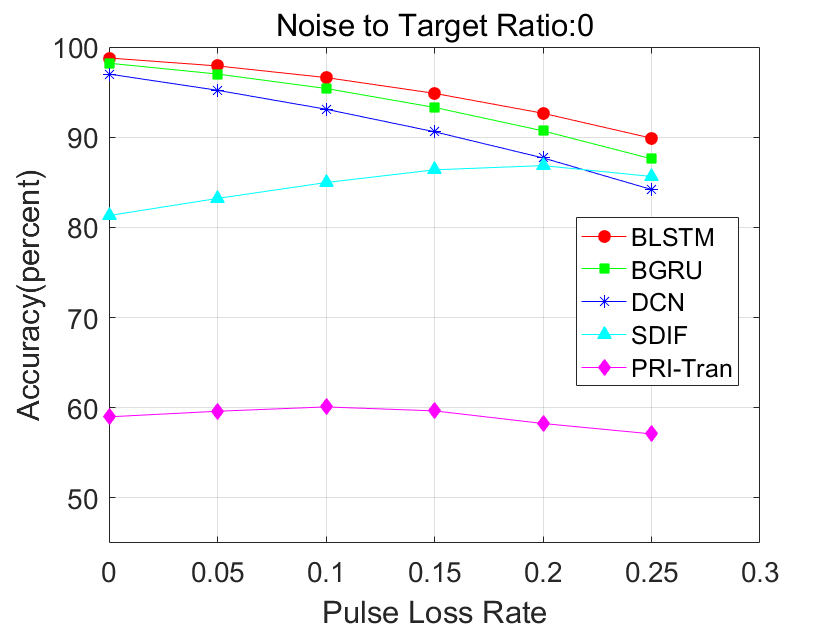}%
\label{exp1a}}
\hfil
\subfloat[]{\includegraphics[width=2.3in]{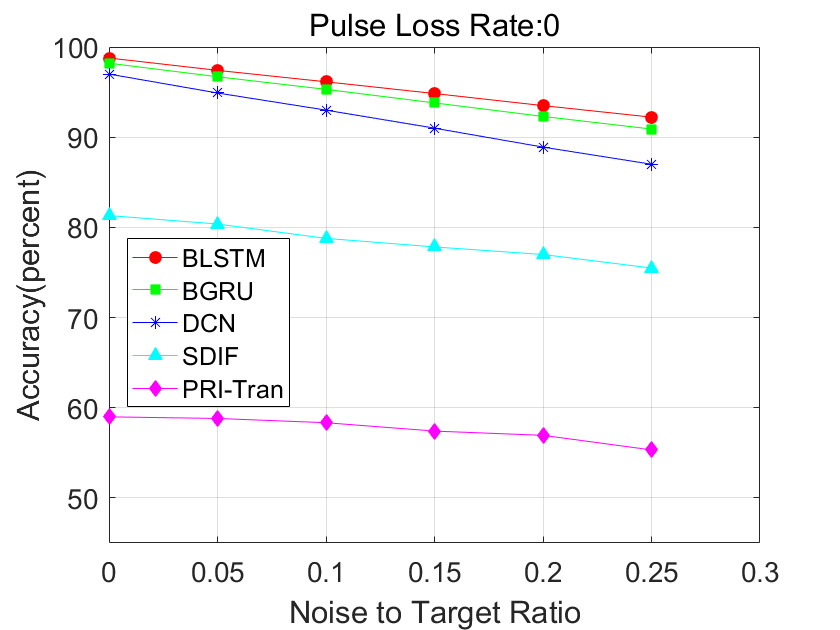}%
\label{exp1b}}
\hfil
\subfloat[]{\includegraphics[width=2.3in]{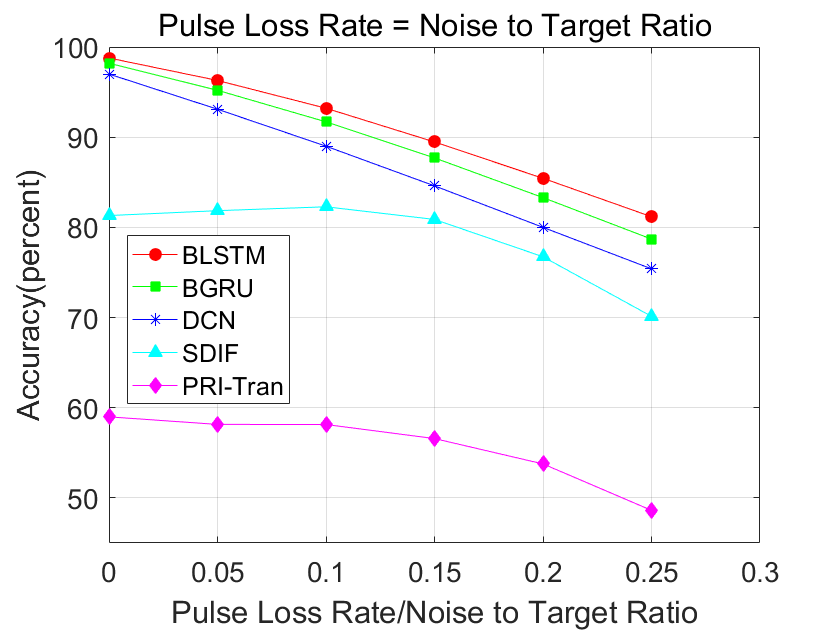}%
\label{exp1c}}
\caption{Deinterleaving radar signal with PRI modulation modes.}
\label{exp1}
\end{figure*}

\begin{figure*}[!t]
\centering
\subfloat[]{\includegraphics[width=2.3in]{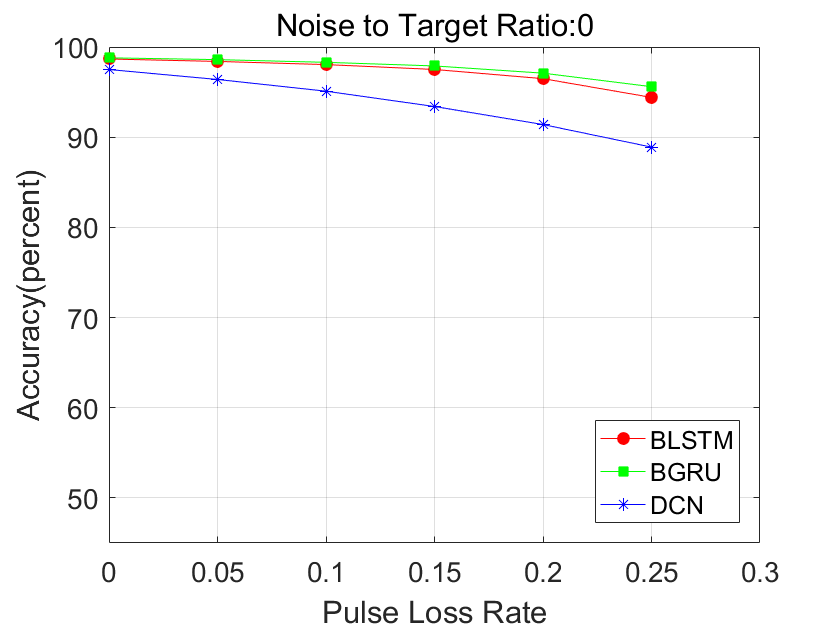}%
\label{exp5a}}
\hfil
\subfloat[]{\includegraphics[width=2.3in]{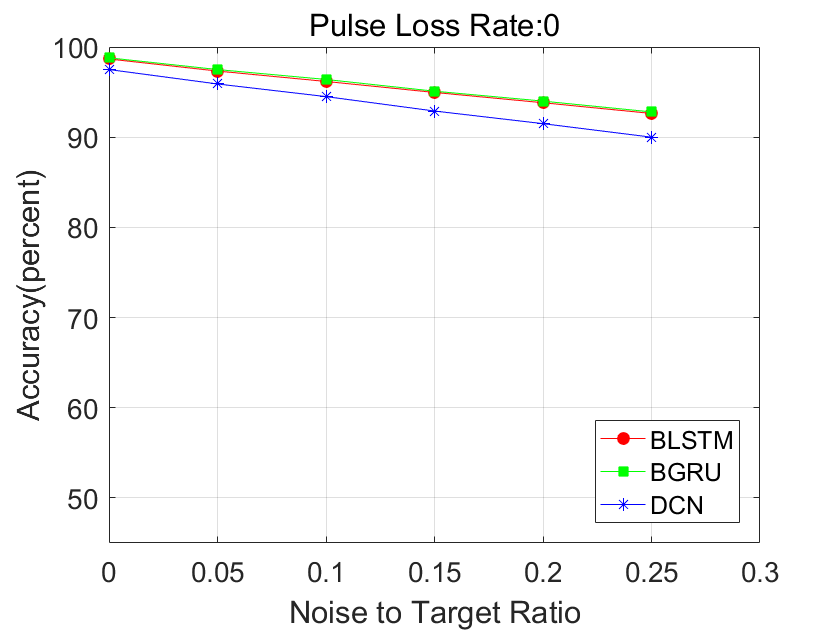}%
\label{exp5b}}
\hfil
\subfloat[]{\includegraphics[width=2.3in]{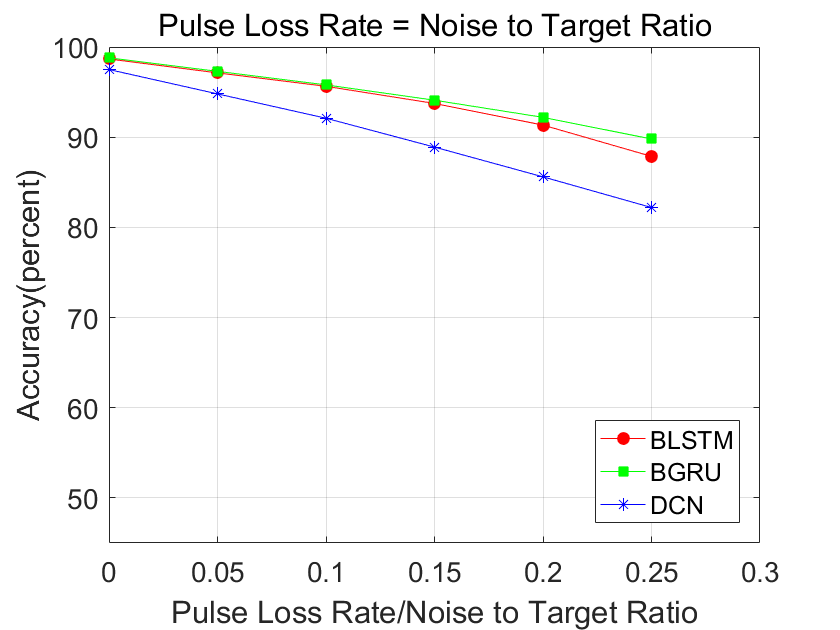}%
\label{exp5c}}
\caption{Deinterleaving radar signal with PRI modulation modes
when there are multiple targets per PRI modulation mode.}
\label{exp5}
\end{figure*}

In addition, we tested the trained model on datasets generated under three different conditions: a) there is pulse loss, but no random noise pulse, i.e., ${{\rho }_{n}=0}$; b) there is no pulse loss, i.e., ${{\rho }_{l}=0}$, but there are random noise pulses; c) there are both pulse loss and random noise pulses, and ${{\rho }_{l}={\rho }_{n}}$. The results are shown in Figs. \ref{exp1}—\ref{exp5}.

The experiments have proven the feasibility of the radar signal deinterleaving method based on semantic segmentation and the deinterleaving strategy proposed in Section IV. This method does not require the target PRI to be found first and can adapt to complex PRI modulation modes with superior accuracy and robustness. The experiments also proves that the classical RNN architecture is better than the CNN architecture in this task.

From the results of controlled experiments, SDIF and PRI-TRAN  methods often do not achieve the best deinterleaving accuracy when the data quality is the best. This is because we adjust the threshold to make the overall performance of these methods the best in the sample, which also reflects the shortcomings of such methods. In addition, in this kind of method, the remaining pulses after sorting are regarded as noise pulses. Therefore, in some cases, when the proportion of noise pulses increases, the overall sorting accuracy will increase, but in fact, the sorting accuracy of target pulses decreases. 

\begin{table}[]
\centering
\caption{Network capacity and overall performance in the five experiments}
\label{capacity}
\resizebox{3in}{!}{
\begin{tabular}{|c|c|ccccc|}
\hline
\multirow{2}{*}{\textbf{experiment}} &
  \multirow{2}{*}{\textbf{model size}} &
  \multicolumn{5}{c|}{\textbf{accuracy}} \\ \cline{3-7} 
 &
   &
  \multicolumn{1}{c|}{\textbf{BLSTM}} &
  \multicolumn{1}{c|}{\textbf{BGRU}} &
  \multicolumn{1}{c|}{\textbf{DCN}} &
  \multicolumn{1}{c|}{\textbf{SDIF}} &
  \textbf{PRI-Tran} \\ \hline
1 &
  $\approx$ 611K &
  \multicolumn{1}{c|}{91.1} &
  \multicolumn{1}{c|}{89.3} &
  \multicolumn{1}{c|}{86.4} &
  \multicolumn{1}{c|}{83.2} &
  60.6 \\ \hline
2 &
  $\approx$ 211K &
  \multicolumn{1}{c|}{98.2} &
  \multicolumn{1}{c|}{94.1} &
  \multicolumn{1}{c|}{86} &
  \multicolumn{1}{c|}{92} &
  86.5 \\ \hline
3 &
  $\approx$ 611K &
  \multicolumn{1}{c|}{95.5} &
  \multicolumn{1}{c|}{95.8} &
  \multicolumn{1}{c|}{71.5} &
  \multicolumn{1}{c|}{83.2} &
  41 \\ \hline
4 &
  $\approx$ 611K &
  \multicolumn{1}{c|}{91.5} &
  \multicolumn{1}{c|}{90.7} &
  \multicolumn{1}{c|}{84.2} &
  \multicolumn{1}{c|}{90.1} &
  79.9 \\ \hline
5 &
  $\approx$ 611K &
  \multicolumn{1}{c|}{95.5} &
  \multicolumn{1}{c|}{95.4} &
  \multicolumn{1}{c|}{83.5} &
  \multicolumn{1}{c|}{\textbackslash{}} &
  \textbackslash{} \\ \hline
\end{tabular}
}
\end{table}

\section{CONLUSION}

In this paper, a radar signal deinterleaving method based on semantic segmentation is proposed. It uses semantic information contained in different radar signals to label pulses that constitute the same semantics as the same category. Two deinterleaving strategies comprehensively using PRI modulation modes and parameters are also proposed. Based on this research, we obtained the following conclusions.

1) Compared to the traditional methods \cite{1993-Wiley}, 
\cite{1989-CDIF}, 
\cite{1992-SDIF}, 
\cite{2000-PRITran}, 
\cite{1999-Spectrum-Estimation-of-Interleaved-Pulse-Trains}, 
\cite{2002-New-Matrix-Method}, 
\cite{1998-A-Novel-Pulse-TOA-Analysis-Technique}, 
\cite{1994-Kalman}, 
\cite{1998-Kalman}, 
\cite{1999-Kalman}, 
\cite{2010-Kalman-MHT},  
\cite{2005-HiddenMarkov}, 
\cite{2019-Improved-Deinterleaving-on-Correlation}, the SSD method does not need to find PRI or PRI period first and then conduct a sequence search, so it can adapt to complex PRI modulation environments. It does not need to iterate over data. For radar signals with multiple pulses in one period, multiple rounds of search and merging operations are not required. After deinterleaving with the SSD method, the PRI modulation mode of target radar signal is known, and PRI modulation mode recognition is no longer needed.

2) Compared to other deinterleaving methods based on NNs and automata 
\cite{2020-lzm-Classification-Denoising}, 
\cite{2020-Denoising-Autoencoders}, 
\cite{2020-Deinterleaving-Autoencoders}, 
\cite{2020-Deinterleaving-Automata}, the SSD method does not require digital data processing, avoiding the resolution problem; this method can output multiple targets in one step with one network, without training a network for each target category and without iterating input and output data repeatedly; only the PRI modulation modes and parameter range are required for training, without very accurate PRI values as prior information in this method.

3) The SSD method is easy to train and converge and still maintains ideal accuracy and good robustness in complex deinterleaving environments with a high pulse loss rate and noise to target ratio.

4) We propose two deinterleaving strategies that comprehensively use PRI modulation modes and parameters. In the future, we will investigate which method is more effective and which scenarios they are applicable to.

5) In this paper, PRI modulation modes and parameters are proposed as semantic information. However, the method proposed in this paper still has limitations when PRI modulation modes of targets are the same and the PRI values or value range overlap. At this point, other parameters of full pulse data such as RF and PW can be used as semantic information to further divide the target into more categories. The deinterleaving method based on semantic segmentation and multi-parameter will also be our future research direction.

6) Our research shows that the LSTM and GRU have obvious advantages over DCN in this task. This may indicate that, although the performance of the CNN architecture has exceeded that of the RNN architecture in representative sequence modeling tasks 
\cite{2018-TCN-RNN}, 
RNNs are still superior to CNNs in sequences with strong mathematical relationships. This problem depends on further research and proof in the field of deep learning in the future.

\begin{figure*}[!t]
\centering
\subfloat[]{\includegraphics[width=2.3in]{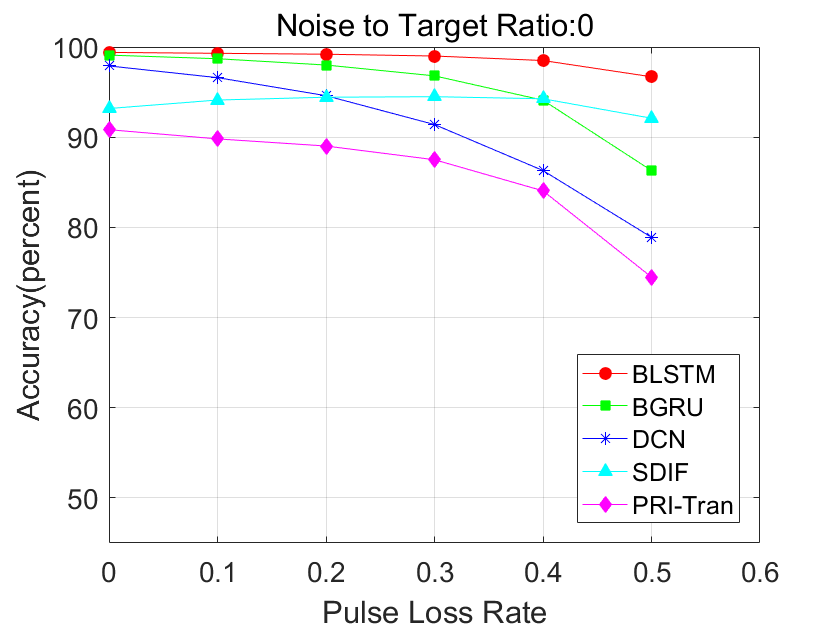}%
\label{exp2a}}
\hfil
\subfloat[]{\includegraphics[width=2.3in]{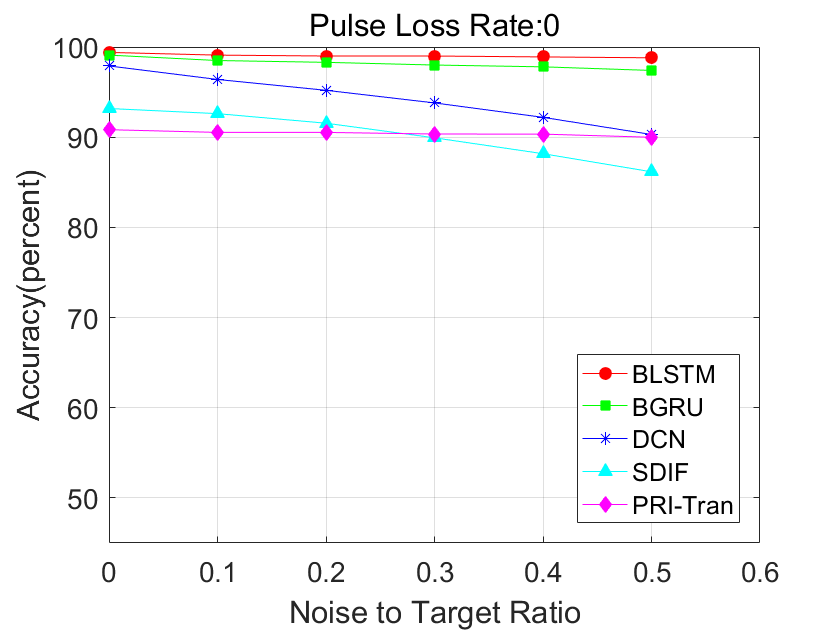}%
\label{exp2b}}
\hfil
\subfloat[]{\includegraphics[width=2.3in]{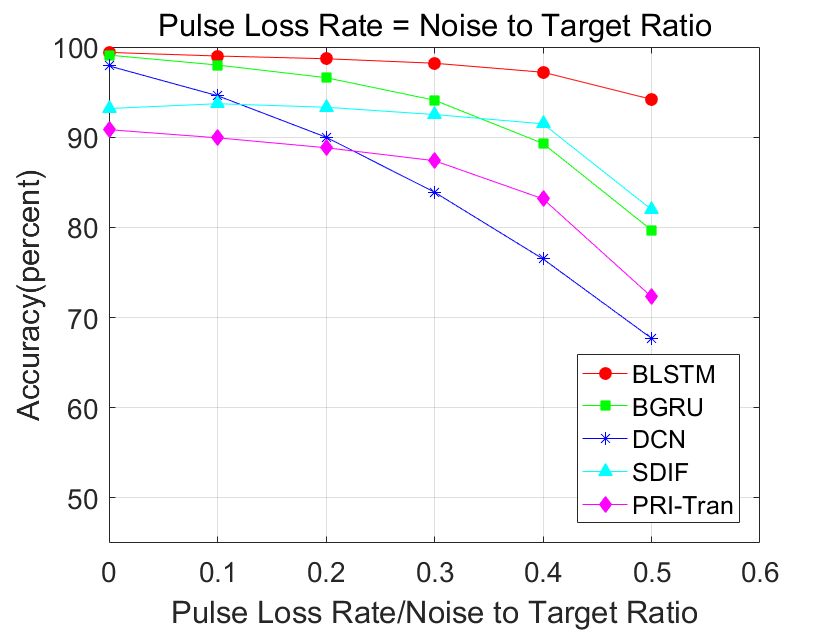}%
\label{exp2c}}
\caption{Deinterleaving radar signal of constant PRI with PRI
values.}
\label{exp2}
\end{figure*}

\begin{figure*}[!t]
\centering
\subfloat[]{\includegraphics[width=2.3in]{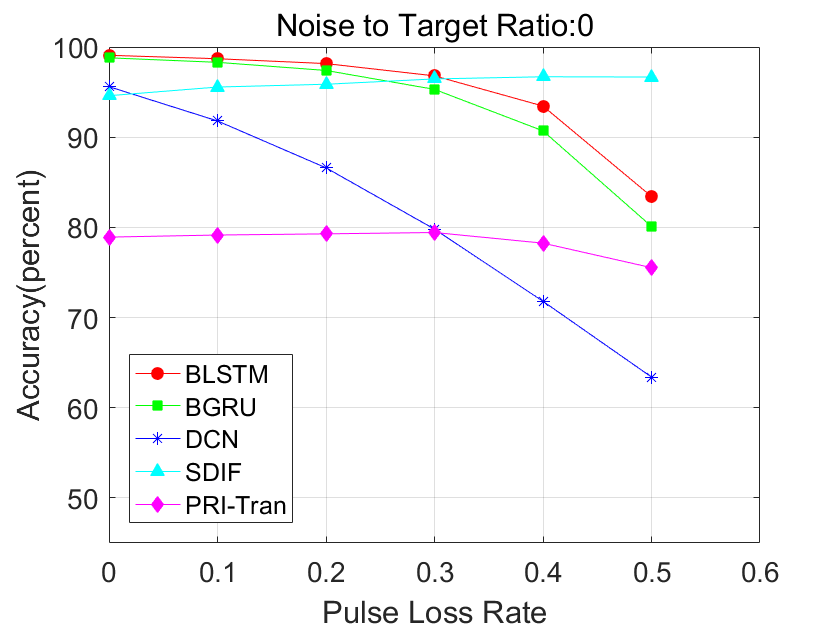}%
\label{exp3a}}
\hfil
\subfloat[]{\includegraphics[width=2.3in]{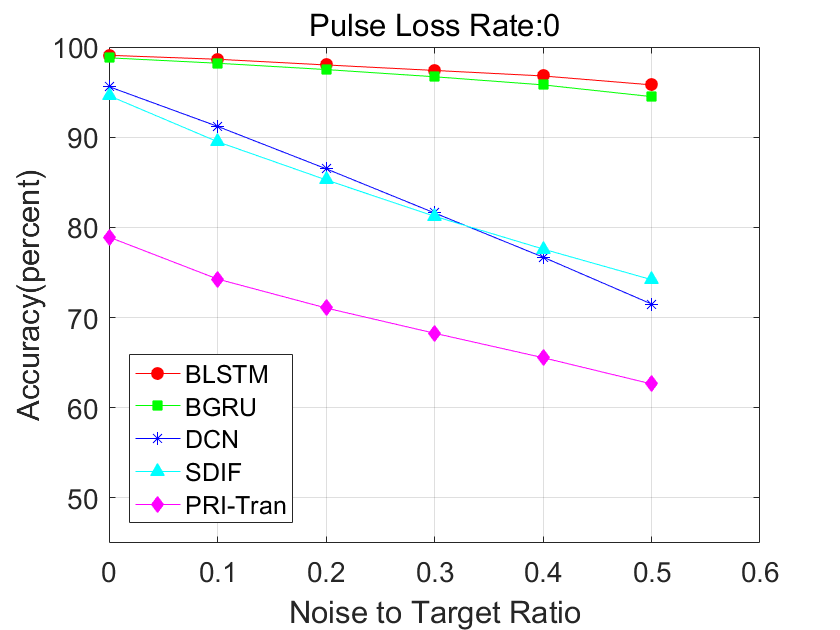}%
\label{exp3b}}
\hfil
\subfloat[]{\includegraphics[width=2.3in]{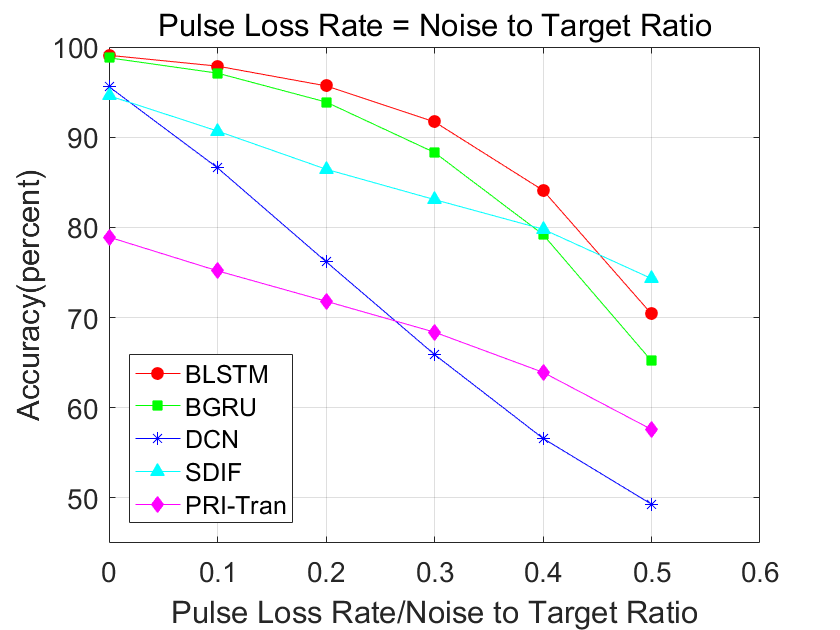}%
\label{exp3c}}
\caption{Deinterleaving radar signal of staggered PRI with PRI
values.}
\label{exp3}
\end{figure*}

\begin{figure*}[!t]
\centering
\subfloat[]{\includegraphics[width=2.3in]{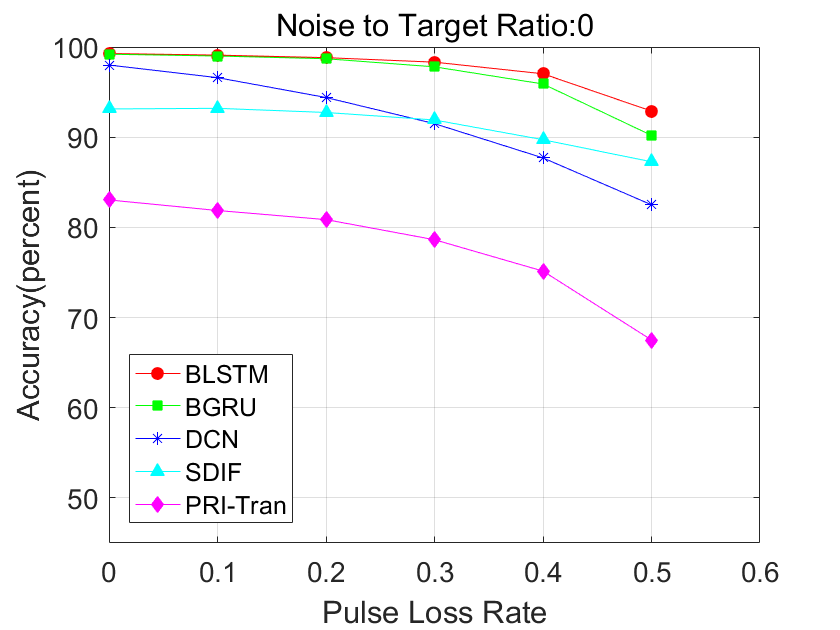}%
\label{exp4a}}
\hfil
\subfloat[]{\includegraphics[width=2.3in]{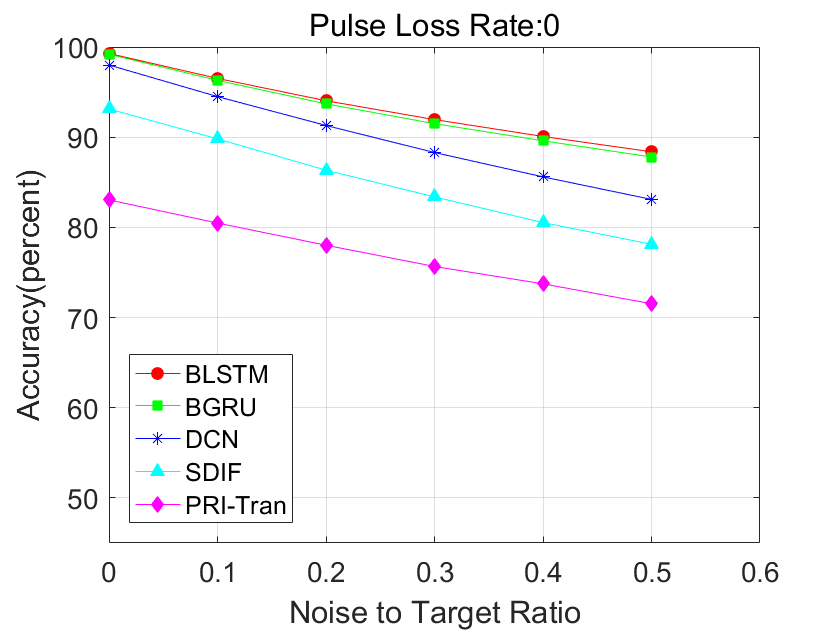}%
\label{exp4b}}
\hfil
\subfloat[]{\includegraphics[width=2.3in]{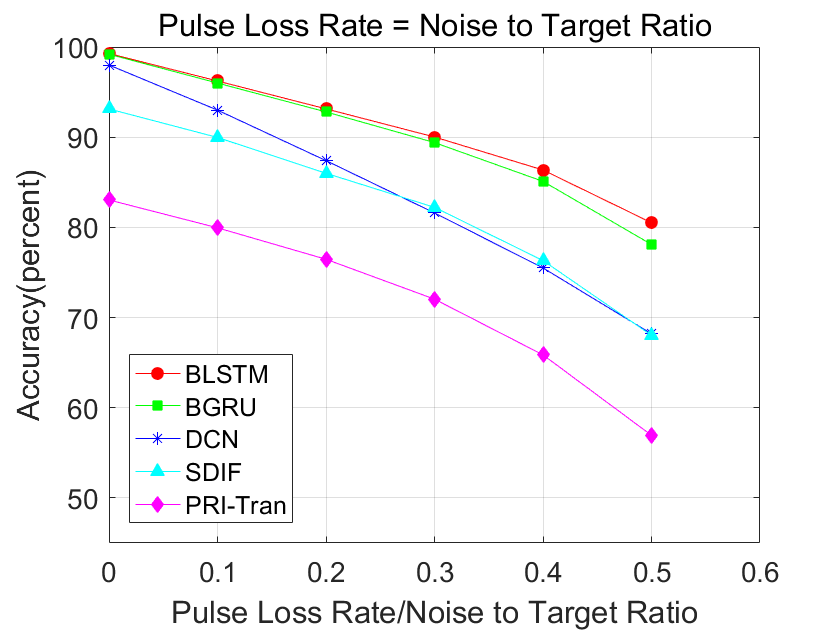}%
\label{exp4c}}
\caption{Deinterleaving radar signal using PRI modulation modes
and PRI parameters simultaneously.}
\label{exp4}
\end{figure*}

\bibliography{Library}

% Can use something like this to put references on a page
% by themselves when using endfloat and the captionsoff option.
\ifCLASSOPTIONcaptionsoff
  \newpage
\fi

\end{document}